\documentclass[twocolumn]{aastex62}

\usepackage{graphicx}
\usepackage{amssymb}
\usepackage{amsmath}
\usepackage{natbib}
\usepackage{wrapfig}
\usepackage[utf8]{inputenc}
\usepackage[T1]{fontenc}
\usepackage{multirow}
\newcommand{\gaia}{\textit{Gaia}}
\newcommand{\hst}{{\it HST}}

\submitjournal{ApJ}

\shorttitle{SMC from Gaia}
\shortauthors{Zivick et al.}

\begin{document}


\title{Deciphering the Kinematic Structure of the Small Magellanic Cloud through its Red Giant Population}

\correspondingauthor{Paul Zivick}
\email{pjz2cf@virginia.edu}

\author[0000-0001-9409-3911]{Paul Zivick}
\affiliation{Mitchell Institute for Fundamental Physics and Astronomy and Department of Physics and Astronomy, Texas A\&M University, College Station, TX 77843, USA}
\affiliation{Department of Astronomy, University of Virginia, 530 McCormick Road, Charlottesville, VA 22904, USA}

\author{Nitya Kallivayalil}
\affiliation{Department of Astronomy, University of Virginia, 530 McCormick Road, Charlottesville, VA 22904, USA}

\author{Roeland P. van der Marel}
\affiliation{Space Telescope Science Institute, 3700 San Martin Drive, Baltimore, MD 21218, USA}
\affiliation{Center for Astrophysical Sciences, Department of Physics \& Astronomy, Johns Hopkins University, Baltimore, MD 21218, USA}

\begin{abstract}

We present a new kinematic model for the Small Magellanic Cloud (SMC), using data from the \gaia\ Data Release 2 catalog. 
We identify a sample of astrometrically well-behaved red giant (RG) stars belonging to the SMC and cross-match with publicly available radial velocity (RV) catalogs.
We create a 3D spatial model for the RGs, using RR Lyrae for distance distributions, and apply kinematic models with varying rotation properties and a novel tidal expansion prescription to generate mock proper motion (PM) catalogs.
When we compare this series of mock catalogs to the observed RG data, we find a combination of moderate rotation (with a magnitude of $\sim10-20$ km s$^{-1}$ at 1 kpc from the SMC center, inclination between $\sim50-80$ degrees, and a predominantly north-to-south line of nodes position angle of $\sim180$ degrees) and tidal expansion (with a scaling of $\sim10$ km s$^{-1}$ kpc$^{-1}$) is required to explain the PM signatures. 
The exact best-fit parameters depend somewhat on whether we assess only the PMs or include the RVs as a qualitative check, leaving some small tension remaining between the PM and RV conclusions.
In either case, the parameter space preferred by our model is different both from previously inferred rotational geometries, including from the SMC H{\small I} gas and from the RG RV-only analyses, and new SMC PM analyses which conclude that a rotation signature is not detectable.
Taken together this underscores the need to treat the SMC as a series of different populations with distinct kinematics.



\end{abstract}

\section{Introduction}\label{sec:intro}

More than a decade ago, our paradigm for the Magellanic Clouds shifted with the measurement of proper motions for the Large and Small Magellanic Clouds (LMC and SMC, respectively). 
Proper motion (PM) measurements of the Clouds made using the \textit{Hubble Space Telescope} (\hst) \citep{NK06b, NK06a} in conjunction with orbital integrations of their interactions with the Milky Way (MW) revealed a strong preference for the first infall scenario \citep{besla07}, supported by follow-up PM analysis shortly thereafter \citep{piatek08}. 
Since then, both the systemic properties and internal dynamics of the LMC have been well-studied \citep[e.g.,][]{olsen11, NK13, vdM14, platais15}. 
However, its smaller companion, the SMC, has proven significantly more challenging to fully characterize, in large part due to its complicated interaction history with the LMC.

In particular, the question of rotation in the SMC has remained frustratingly inconclusive. 
\cite{stanimirovic} produced one of the first detections of a gradient in the radial velocities (RVs) of the H{\small I} in the SMC, finding that the gradient was well fit by a rotating disk, up to a velocity of 60 km s$^{-1}$, inclined out of the plane of the sky. 
An examination of RVs of the red giant (RG) population of the SMC by \cite{harris06} did not reproduce this, instead finding inconclusive evidence for rotation in the SMC. 
Interestingly, a RV study of the OBA stars in the SMC \citep{evans08} did find evidence for rotation in the stars ($\sim 26$ km s$^{-1}$ deg$^{-1}$), albeit with a different inclination angle and line-of-nodes position angle (LON PA) than those found for the H{\small I}, creating a complicated picture of rotation in the SMC.
Two more recent studies have reinforced this disagreement between the type of tracer and the resulting inferred rotation.
\cite{dobbie14a} conducted a spatially broader study of the RG stars than \cite{harris06}. 
With the larger sample, they found a measurable gradient in the RVs, inferring an observed rotation curve between 20-40 km s$^{-1}$, potentially larger than their measurement of the internal dispersion of the SMC ($\sim$ 26 km s$^{-1}$), and an inclination and LON PA consistent with that of \cite{evans08}. 

For H{\small I}, \cite{diteodoro19} provided the highest resolution measurements of the gas in the SMC to date. 
Using the new measurements, they found a plane of rotation consistent with that originally inferred by \cite{stanimirovic}, with improved errors and constraints on the uncertainty in the model fit. 
Intriguingly, a slight preference for this rotational geometry was also found in an RV-only analysis of OB stars by \cite{lamb16},  at odds with other stellar measurements. 
However, for all of these works a key uncertainty lies in only measuring motion in one direction, leaving open the possibility for PM measurements (that complete the 3D velocity vector) to potentially begin to resolve these tensions.

The first PM insights about the internal kinematics of the SMC began with \cite{NK06a}, where they measured PMs for five different fields in the inner regions of the SMC with \hst,\ enabling a measurement of the center-of-mass motion, but not strong constraints on the internal velocity field. 
\cite{NK13} improved the precision for these five fields, but the spatial distribution of the fields prevented further analysis into the question of rotation. 
\cite{vdM16} used PMs from the Tycho-\textit{Gaia} Astrometric Solution Catalog \citep{lindegren16}, a combination of \gaia\ Data Release 1 \citep{gaiadr1a, gaiadr1b} and the \textit{Hipparcos} Tycho-2 Catalog \citep{hoeg00}, to add another eight PM data points to constrain the velocity field of the SMC. 
Their resulting analysis found inconclusive evidence for rotation, limited in part by the relatively small number of data points. 

\cite{zivick18} attempted to expand this data set, using \hst\ to measure PMs for another 28 fields, sampling a broader region around the SMC, but kept their analysis for rotation restricted to the plane of the sky as well. 
While the measurement of a rotation signal proved inconclusive, the analysis did reveal coherent motions in the southeastern portion of the SMC moving radially outwards, in the direction of the Magellanic Bridge and LMC, potentially indicative of tidal expansion. 
Finally, the release of \gaia\ Data Release 2 (DR2) \citep{gaiadr218b} vastly expanded the catalog of PMs for the SMC, opening up for the first time the ability for the PM data to constrain the inclination and LON PA of the rotation model in the SMC. 

The initial analysis of the SMC \citep{gaiahelmi18} found a weak rotation signal throughout the SMC ($\sim8$ km s$^{-1}$ at 3 kpc from SMC center) at a large inclination angle and a LON PA different from that measured from the RG stars in \cite{dobbie14a} and the H{\small I} in \cite{diteodoro19}. 
This departure from RV-inferred parameters continued with an analysis of the young, massive stars (which should trace the H{\small I} kinematics) in \cite{murray19}, where they combined \gaia\ PMs with RVs from spectroscopic surveys and found kinematics inconsistent with coherent disk rotation.
\cite{deleo20} also found kinematics inconsistent with rotation, this time in the RG population, using \gaia\ PMs and an expanded RV catalog for the RGs, but they note evidence for significant tidal fragmentation in the RG population.
\cite{grady20} also observed this tidal stripping in the RGs as well, finding trends not only in PM space from the SMC towards the LMC, but in metallicity space as well.

Given the unusual nature of the SMC and its interaction history with the LMC, constraining the presence of rotation within the SMC could provide us a window into its original form. 
The magnitude of rotation and its orientation, contrasted against the internal dispersion, may help us to understand the SMC's progress in possibly transitioning from a dwarf irregular to a dwarf spheroidal galaxy \citep{besla12}.
Further, these constraints on rotation in turn become a new requirement for any simulations between the LMC and SMC, furthering our understanding of both the Magellanic system and potentially their joint history with the MW.

In simulations we have perfect knowledge of the 6D phase space information of the motions and locations of particles, but creating a  comparable observational data set is challenging. 
\gaia\ DR2, however, now provides us an opportunity to synthesize all the observational efforts aimed at understanding SMC structure and to begin to piece together a more holistic picture of the SMC.
Here, we use a forward modeling approach,  creating a mock SMC data set that we transform into observational kinematic space, with a view to match the \gaia\ data. 
In addition to the tens of thousands of PMs in the \gaia\ DR2 Catalog, we cross-match with publicly available RV studies, bringing us closer to producing a truly 3D observational data set for the stars. 

In Section \ref{sec:data}, we describe our selection of \gaia\ DR2 data, both for astrometric quality and SMC membership, as well as cross-matching with existing RV catalogs. 
Section \ref{sec:model} first details the creation of our spatial model of the SMC and the kinematic model applied to create a mock data catalog. 
We then outline our methodology for assessing the fit of a given model to the data. 
In Section \ref{sec:comp} we discuss the parameter space explored for our models and present the best fitting class of models.
Finally, in Section \ref{sec:conc} we summarize our efforts in modeling the SMC and forecast ahead to future work for further improving our understanding of the SMC internal kinematics.

\section{Data Selection}\label{sec:data}

\subsection{Gaia DR2 Selection}\label{ssec:gaiadata}

\begin{figure}
\centering
 \includegraphics[width=3.3in]{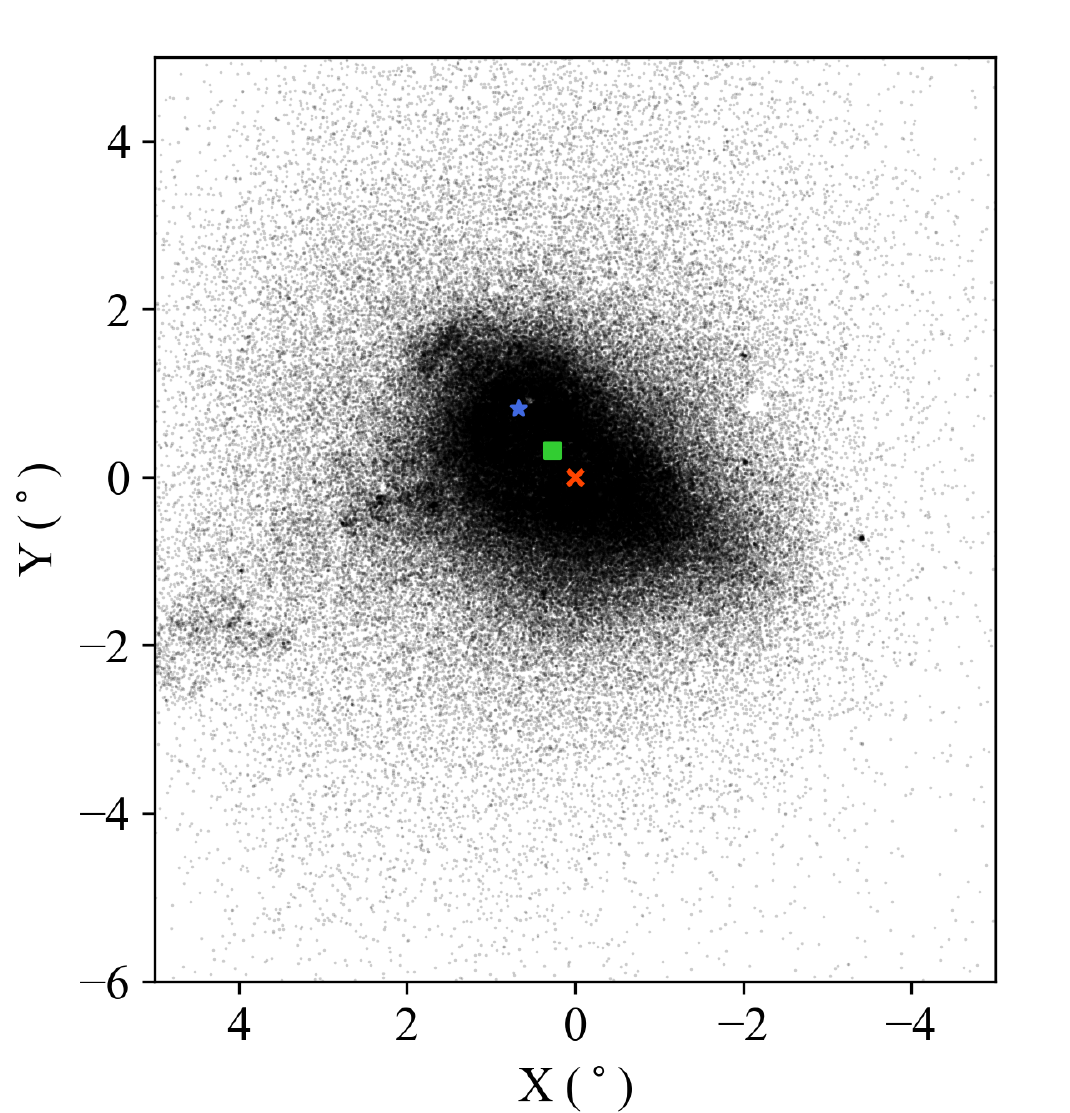} 
 \caption{\gaia\ DR2 sources after initial astrometric cuts have been applied to the data, displayed in a Cartesian coordinate system as outlined in \cite{gaiahelmi18}. 
 The zero point is set to the best-fit center for the RG population (the red ``x''), described in Section \ref{sec:data}. 
 The inferred kinematic center from H {\small I} gas is marked by the blue star, and the geometric center derived from the RR Lyrae population is marked by the green square. 
 The SMC Wing can be seen on the eastern (left) side of the SMC, heading towards the Magellanic Bridge and the LMC. 
 We also note the empty patch on the right side of the figure as the location of 47 Tuc, which has been removed from our sample by our astrometric criteria.}
   \label{f:mc}
\end{figure}

For our analysis, we select all stars from the \gaia\ DR2 catalog within roughly 5 degrees from the optical center of the SMC using \texttt{pygacs}\footnote{\small \url{https://github.com/Johannes-Sahlmann/pygacs}}. 
We then apply a series of initial cuts to jointly select both astrometrically well-behaved stars and stars likely to belong to the SMC. 
To remove MW foreground stars, we require all stars in our sample to have a parallax $< 0.2$ mas and a proper motion within 3 mas yr$^{-1}$ of the SMC systemic motion
($\mu_{W} = -0.82 \pm 0.1$ mas yr$^{-1}$ and $\mu_{N} = -1.21 \pm 0.03$ mas yr$^{-1}$ from \cite{zivick18}, where $\mu_{W} \equiv -\mu_{\alpha}\mathrm{cos}\delta$ and $\mu_{N} \equiv \mu_{\delta}$ in \gaia-provided quantities).

Next we apply the following cut to the renormalized unit weight error (RUWE) as described in the \gaia\ technical note GAIA-C3-TN-LU-LL-124-01:
\begin{equation}
    \frac{\sqrt{\chi^2 / (N-5)}} {u_{0}(G,C)} < 1.40,
\end{equation}
which uses the following \gaia\ properties:
\begin{equation}
\begin{split}
    N \equiv \texttt{astrometric\_n\_good\_obs\_al}, \\
    \chi^2 \equiv \texttt{astrometric\_chi2\_al}, \\
    u_{0} \equiv \mathrm{Normalization\ factor}\left(G, C\right), \\
    G \equiv \texttt{phot\_g\_mean\_mag}, \\
    C \equiv \texttt{bp\_rp}. 
\end{split}
\end{equation}
 
We additionally apply a cut for the color excess of the stars, as described in \cite{gaiadr2hrd} Equation C.2. 
As we are concerned primarily with the bright, astrometrically well-behaved stars, and to provide another check to avoid MW contamination, we select stars brighter than $G < 18$, leading to the final source densities in Figure \ref{f:mc}. 
We note the conspicuous absence of 47 Tuc on the right side of Figure \ref{f:mc} as an example of the power of \gaia \ DR2 to remove potential contamination from spatially coincident sources.

\begin{figure}
\centering
 \includegraphics[width=3.3in]{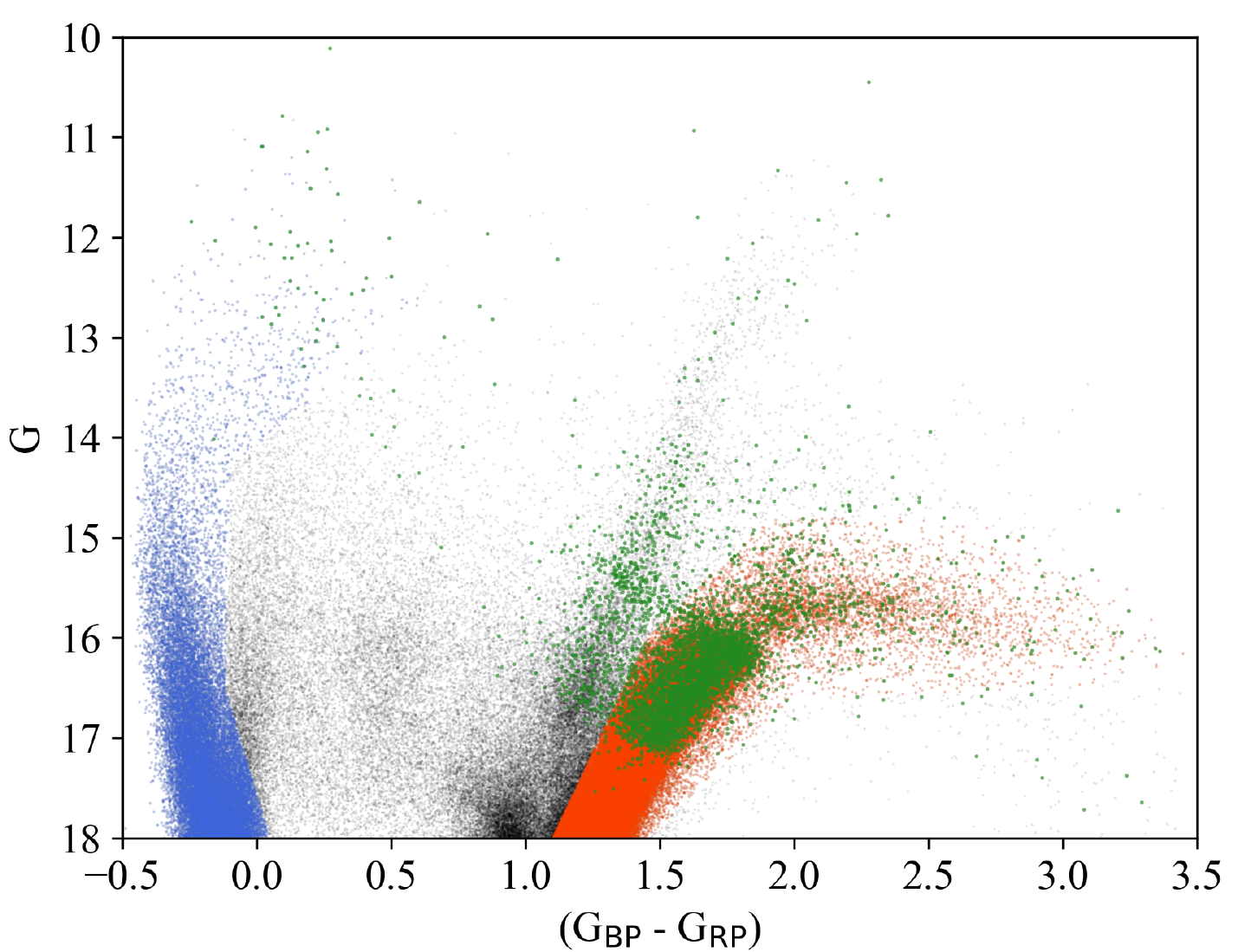} 
 \caption{Color-Magnitude Diagram of all sources present in Figure \ref{f:mc}. 
 Two separate stellar sequences have been marked: the main sequence (MS) stars in blue on the left side and the red giants (RG) on the right side. 
 The top of the red clump can be observed near (1.0, 18.0), but due to worsening astrometric performance near G $\sim18$, we have chosen not to examine it further. 
 All sources that have been matched to an existing RV measurement (described in Section \ref{ssec:rvdata}) are marked in green, which can be seen to mostly sample the RGs but do extend to the sequence blueward of the RGs, likely a red supergiant population in the SMC.}
   \label{f:cmdtypes}
\end{figure}

With this initial astrometric selection we move to further isolate the SMC stars. 
Examining the color-magnitude diagram (CMD) in Figure \ref{f:cmdtypes}, we identify two clear SMC stellar tracks in our sample: the main sequence (MS, marked in blue) and the red giants (RG, red). 
What we have defined as the RG sequence includes both standard RG branch stars as well as a population of carbon stars (the more redward extension at G$\sim$16). 
However, prior work has shown that the kinematics of the carbon stars appears consistent with the standard RG stars \citep{vdM02}, so to improve our statistics, we use both populations as our ``RG'' sample. 
The SMC red clump is also observable near the bottom of the CMD, but due to worsening astrometric performance at G $> 18$, we do not examine it further in this work. 
We also present all sources with a cross-matched RV (see Section \ref{ssec:rvdata} for more details) in green. 

We do note the presence of a likely third stellar sequence, a group of red supergiants (RSG) located just blueward of the RG branch, that contains a number of the RV-measurements (indicated by the purple points overlaid on the CMD). 
Rudimentary isochrone fits to the RSG sequence find a good agreement to an age $\sim150$ Myr. 
As this age is consistent with the most recent time of peak interaction between the LMC and SMC \citep{zivick18, martinez-delgado19, joshi19}, which may introduce complex kinematic signatures into the data, we choose to leave a more thorough study of this intermediate age population to a future effort.

\begin{figure}
\centering
 \includegraphics[width=3.3in]{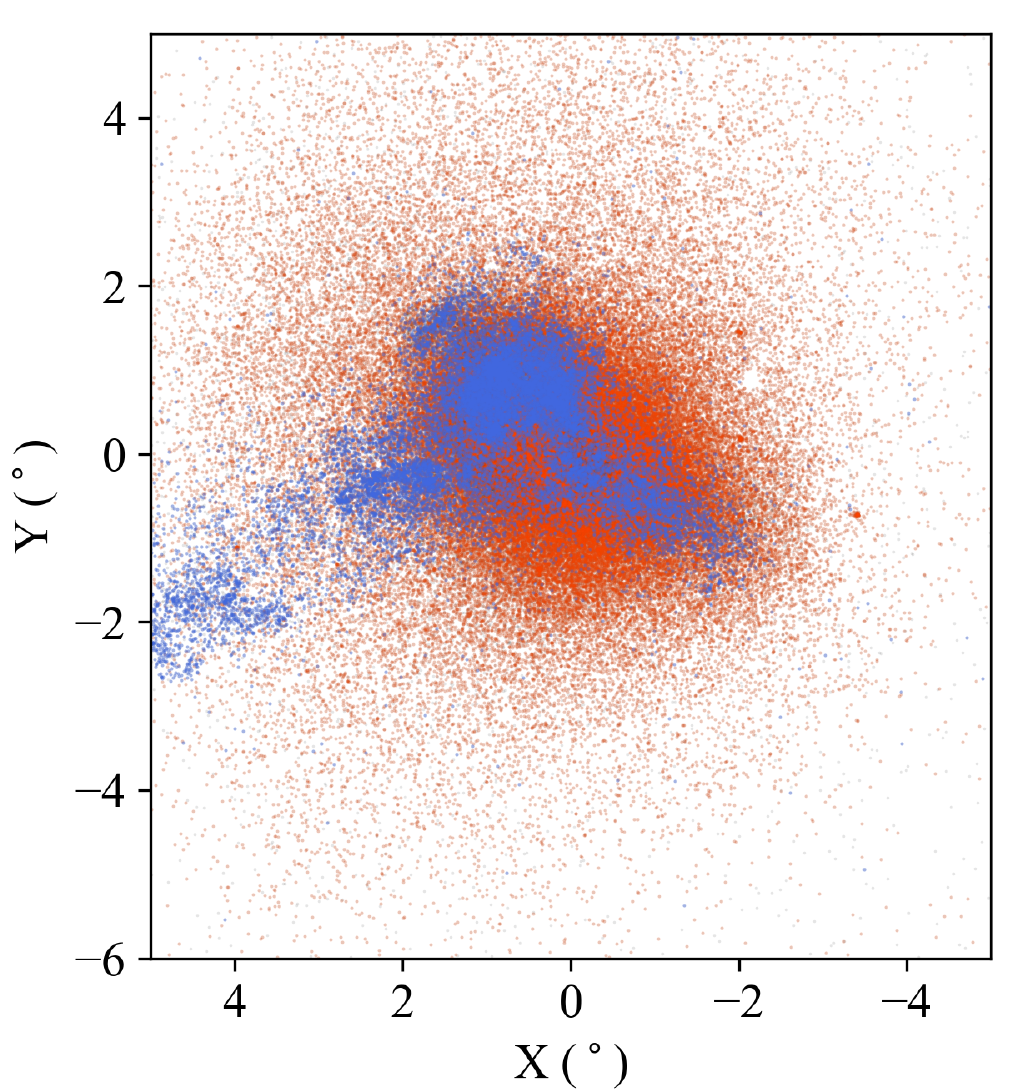}
 \caption{All SMC main sequence (MS, blue) and red giant (RG, red) stars from Figure \ref{f:cmdtypes} in the same field of view and Cartesian coordinate system as Figure \ref{f:mc}. 
 The MS stars have a clear spatial structure to them, tracing the optical main body of the SMC and stretching into the Wing, while the RGs have a significantly more well-behaved structure, appearing to be roughly azimuthally isotropic with a radial dependence for the density. 
 }
   \label{f:bothpos}
\end{figure}

\begin{figure}
\centering
 \includegraphics[width=3.3in]{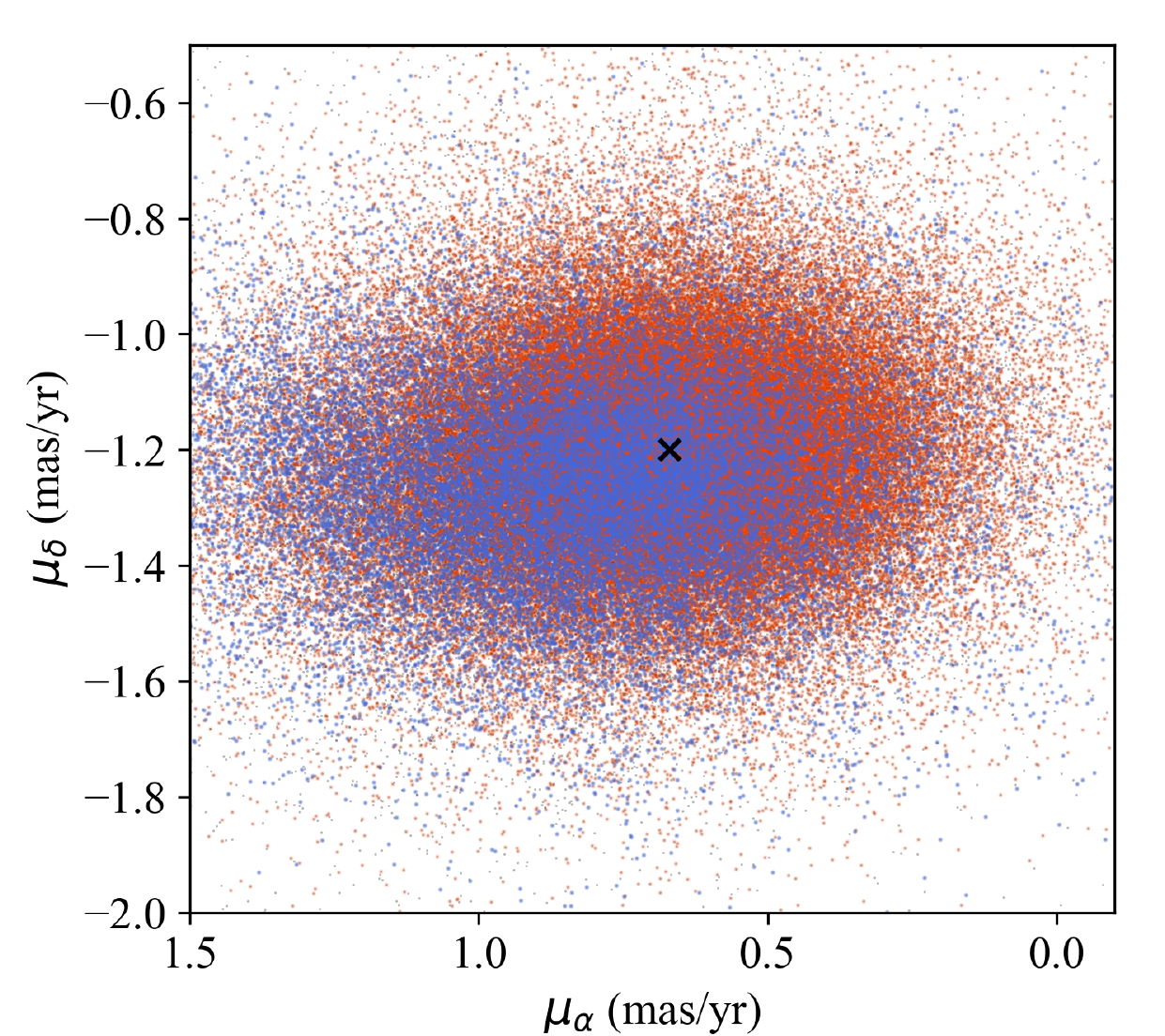}
 \caption{Distribution of the PMs of the RG and MS samples. The PM center of the RGs, measured with a simple Gaussian fit, is marked by the black ``X''. 
 The RG PM is clearly offset from where one would place the MS center, though the asymmetric extension of the MS PMs towards the left side of the figure would make it difficult to accurately assess the systemic MS PM.
 }
   \label{f:bothpm}
\end{figure}

We next consider the spatial structure of these populations in the top panel of Figure~\ref{f:bothpos}. 
The MS and RG populations show a marked difference in their structure. 
The MS stars occupy an irregular distribution, roughly tracing the main optical body of the SMC with minimal presence in the SMC halo, and with numerous smaller clumps embedded within the larger structure. 
The RG population, however, appears to be approximately azimuthally isotropic, with only a radial dependence for its spatial density.
This difference extends into the PM space as well (Figure \ref{f:bothpm}), where the center of the RG PM distribution, fit with a Gaussian, is markedly offset from the bulk of the MS PMs.
The distribution of the PMs also appears different, with the RGs having a smoother distribution while the MS stars have clear asymmetries towards the left (eastward) side of the figure. 
Given the smoother distribution of the RGs in both spatial and kinematic spaces and the wealth of literature on the structure of older stars in the SMC \citep[e.g.,][]{ss_sa12, jacyszyn17}, we choose to only concentrate on the RGs for our current analysis.

\subsection{Radial Velocity Cross-Matching}\label{ssec:rvdata}

To attempt to provide full 3D velocity information for our selected stars, we look to cross-match our catalog with existing RV catalogs.
For this we use two publicly available catalogs: the prior work by \cite{dobbie14a} and RVs from the Apache Point Observatory Galactic Evolution Experiment 2 (APOGEE-2) \citep{Majewski17, Wilson19}, part of the Sloan Digital Sky Survey IV \citep{Blanton17} Data Release 16 \citep{SDSSDR16}. 
Between the two catalogs, excluding any sources that may have been double counted, we find over 4,000 sources common to our DR2 RG selection. 

After a further examination of this RV catalog, we ultimately choose not to include it in our quantitative data-to-model evaluation. 
As will be more fully explained in Section \ref{ssec:modeldes}, the introduction of the tidal expansion component adds a significant distance-dependent effect for a given stellar motion.
In a regime like the \gaia\ PMs with over 100,000 stars, we expect to be sampling the full distance distribution, a necessary requirement for comparing our 3D spatial model to the data.
However, given the comparably sparser sampling of the RV stars, we cannot be sure that requirement is met.
As such, we limit the RV catalog to providing a qualitative check on the consistency and predictions of our models.


\section{Data Analysis}\label{sec:model}

\subsection{Observational Data}\label{ssec:obsdata}

Before any analysis can begin of the internal kinematics of the SMC RGs, we must first deal with correcting for viewing perspective effects, which cause the 3D motion of the SMC to project differently as we change our line of sight. 
However, this correction requires assumptions regarding the location (in distance and on the sky) and magnitude of the 3D motion. 
As \gaia\ affords us the opportunity to only examine individual stellar populations (as opposed to averaging over multiple populations), we derive the systemic properties for the perspective corrections directly from the RG data.

For this, we fit two 2D Gaussians to the spatial and PM distributions of the RGs, finding a center of (($\alpha$, $\delta$) (J2000) = (13.04$^\circ$,$-$73.10$^\circ$)) and a systemic RG PM of (($\mu_{\alpha}$, $\mu_{\delta}$) = (0.67 mas yr$^{-1}$, $-$1.20 mas yr$^{-1}$). 
The spatial center is marked in Figure \ref{f:mc} and the systemic PM is marked in Figure \ref{f:bothpm}. 
The center generally agrees well with previous analyses of the older stellar structure in the SMC \citep[e.g.,]{jacyszyn17}, and the PM value is in line with other \gaia-based measurements of the RG systemic motion \citep[e.g.,]{deleo20}. 
For a distance to the center, we use the estimates from \cite{jacyszyn17} for a distance modulus of (m$-$M) = 18.91.

\begin{figure}
\centering
 \includegraphics[width=3.3in]{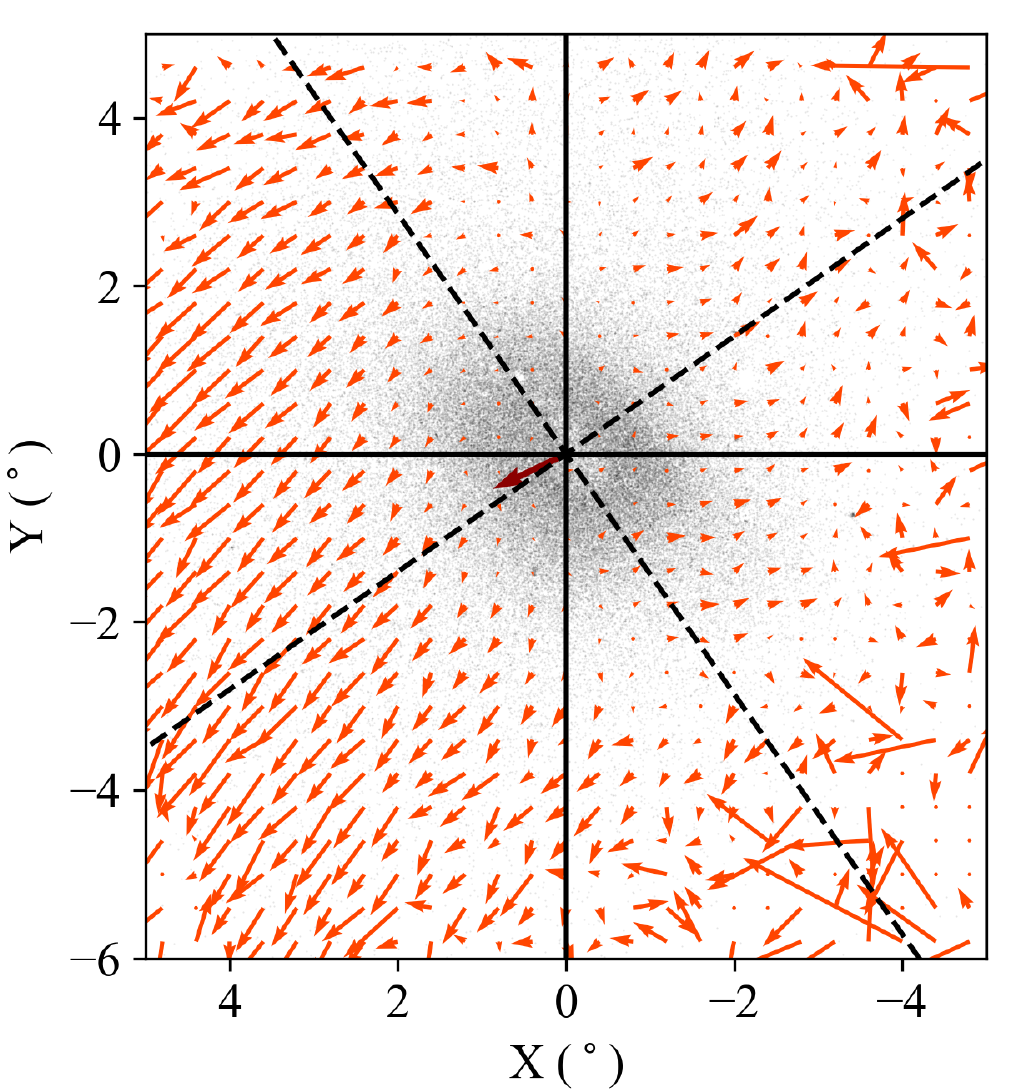}
 \includegraphics[width=3.3in]{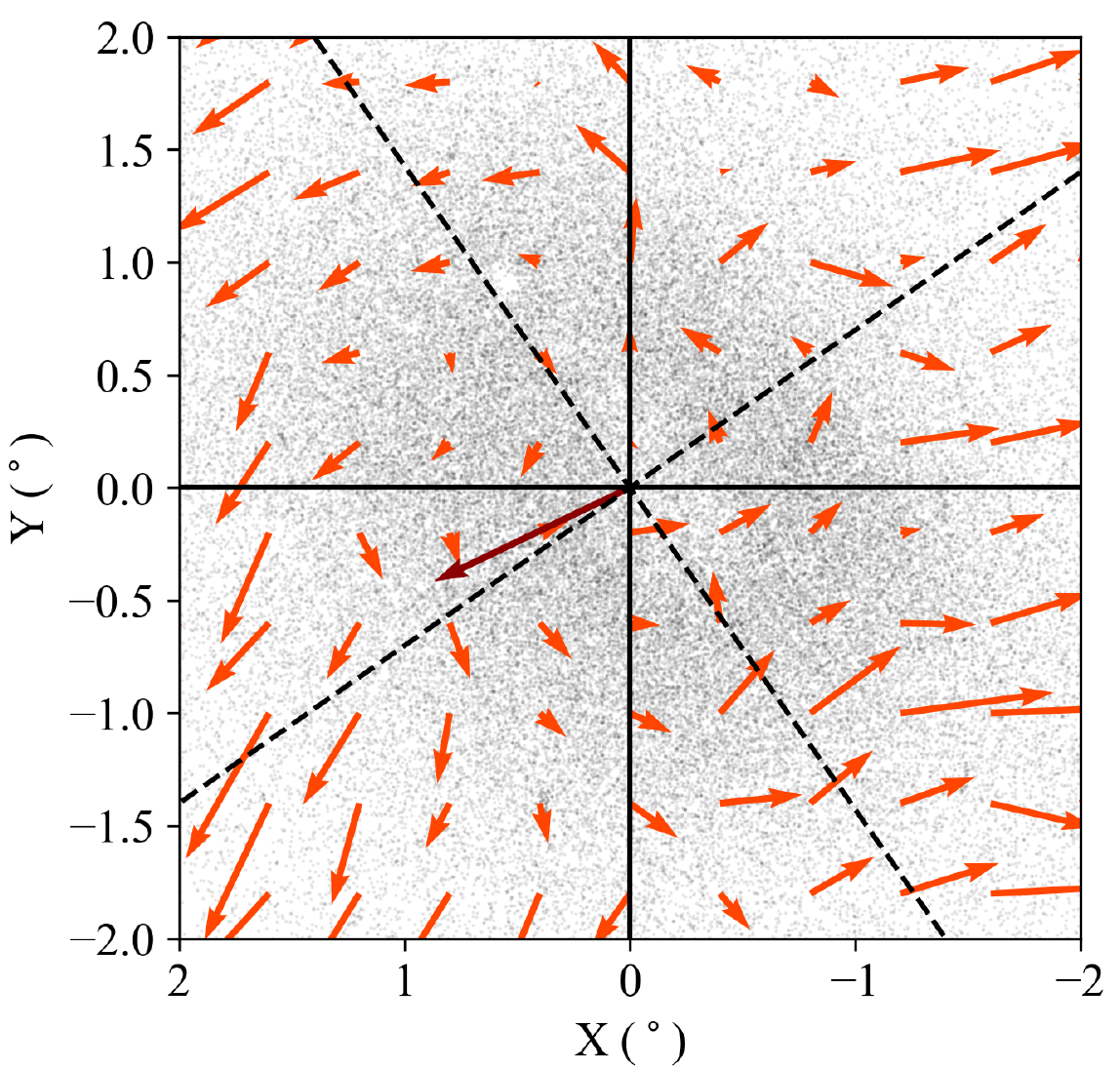}
 \caption{(\textbf{Top}) The average residual PM vectors of the SMC RGs as a function of location in the SMC. The gray points mark all RG stars used in the averaging. 
 The solid black lines indicate the original Cartesian coordinate system, and the dashed black line marks the rotated frame where the $x$-axis lies along the major axis of the SMC RG distribution. 
 The dark red vector marks the relative velocity, as projected on the sky, between the LMC systemic motion and the SMC systemic motion, $\mu_{\mathrm{rel}}$. 
 Clear outward motion can be seen on the eastern side of the SMC, consistent with the direction of $\mu_{\mathrm{rel}}$, while coherent rotation appears to be located in the center of the SMC.
 (\textbf{Bottom}) The inner region of the above plot with the same axes and symbols. However, the vectors have been enhanced by a factor of 6, compared to the above plot, for clarity.
 }
   \label{f:respm_data}
\end{figure}

With these three values in hand, we subtract the systemic PM and perspective correction from each individual PM to calculate the ``residual'' PM. 
We then convert these residual PMs into a Cartesian frame as outlined in \cite{gaiahelmi18}. 
For display, the 100,000 stars are binned every 0.4 degrees and the average residual PM is calculated. 
The resulting averaged residual SMC RG PMs are displayed in Figure \ref{f:respm_data} with the top panel displaying the full SMC window and the bottom panel a zoomed-in plot of the SMC core. 

\begin{figure}
\centering
 \includegraphics[width=3.3in]{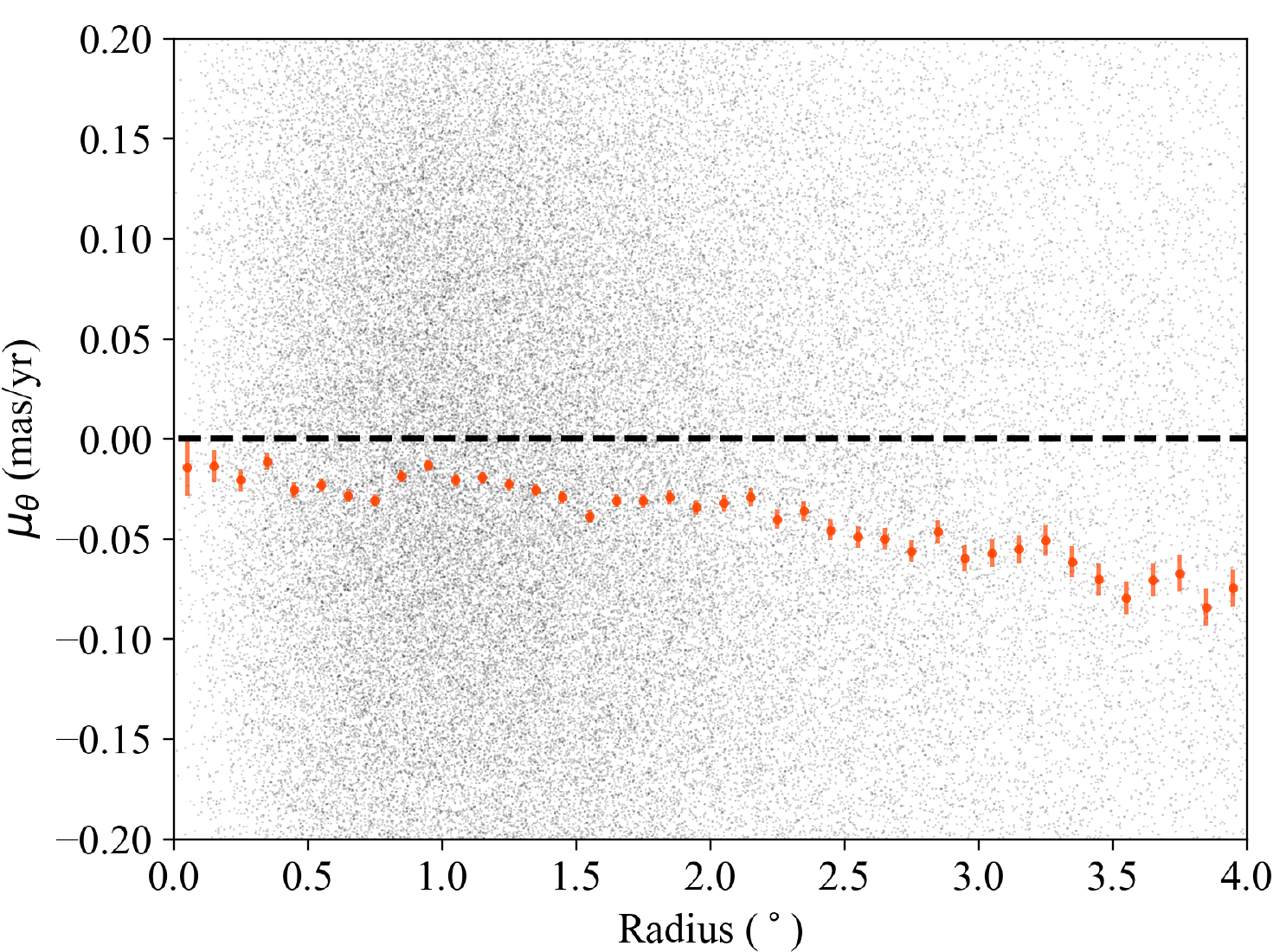}
 \caption{The tangential component of the RG residual PMs, $\mu_{\theta}$, as a function of radius. For easier interpretation, given the large scatter in $\mu_{\theta}$, average values, binned every 0.1 degrees, are marked by the orange circles, with error bars from the error of the mean plotted as well. Across the full radius range, $\mu_{\theta}$ maintains a non-zero average value (with negative $\mu_{\theta}$ corresponding to counter-clockwise rotation), hovering around 0.3 mas yr$^{-1}$ from 0.8 to 2.0 degrees before dramatically increasing.
 }
   \label{f:r_vt_data}
\end{figure}

As other prior analyses have noted, there is a clear outward motion, particularly on the eastern side of the SMC towards the direction of the Magellanic Bridge. 
However, \textit{apparent rotation can be seen in the center of the SMC} with the averaged residual vectors displaying a distinct counter-clockwise pattern in the inner 2 degrees of the SMC. 
To investigate this further, we decompose the residual PM of each star into radial ($\mu_{\mathrm{rad}}$) and tangential ($\mu_{\theta}$) components, relative to the assumed RG center.
Figure \ref{f:r_vt_data} displays $\mu_{\theta}$ (which should be correlated with rotation) as a function of radius for all RG stars.
For easier interpretation, we average $\mu_{\theta}$ every 0.1 degrees along with the calculated error of the mean in Figure \ref{f:r_vt_data}, marked by the orange circles and error bars.
In this space, coherent rotation should appear as a non-zero $\mu_{\theta}$ (above or below zero depending on direction of rotation).
When we examine the data, we see that this is in fact the case, with $\mu_{\theta}$ continuing to rise in magnitude over the full radius, though the peak $\mu_{\theta}$ value (around $\sim0.03$ mas yr$^{-1}$, which corresponds to $\sim9$ km s$^{-1}$ at 60 kpc) stays roughly the same between 0.8 degrees and just past 2 degrees before beginning a significant incline.
While this is not a large value, it does appear to be statistically significant across the full radius.

Returning to the more obvious feature in the outward eastern motions, the idea of tidal expansion or disruption has often been invoked in trying to understand this structure in the residual SMC PMs. 
To place this idea into context, we calculate the relative velocity between the LMC and SMC, $v_{\mathrm{rel}}$. 
We use the new systemic motion for the RG population and take care to perform this subtraction in 3D Galactocentric coordinate space before transforming $v_{\mathrm{rel}}$ back into the observed frame (where we find its components in our Cartesian frame to be ($\mu_{x}$, $\mu_{y}$, $v_{z}$) = (0.43 mas yr$^{-1}$, -0.21 mas yr$^{-1}$, -36.6 km s$^{-1}$)). 
The PM components of $v_{\mathrm{rel}}$, which we will refer to as $\mu_{\mathrm{rel}}$, are displayed as the dark red vector at (0,0) in Figure \ref{f:respm_data}.

Immediately we see that the direction of $\mu_{\mathrm{rel}}$ agrees with the general direction of the residual PMs in addition to the minor axis of the SMC. 
Combined with the hint of coherent rotation in the interior of the SMC (observed in the bottom panel of Figure \ref{f:respm_data}), we propose that the internal velocity field of the SMC can be approximately modeled by two mechanisms: a cylindrical rotation and a linearly increasing tidal component. 
To fully understand this parameter space, we attempt to build a 3D model of the SMC, as described in the following Section.

We do note the potential complications introduced by systematic \gaia\ errors that have a specific spatial frequency, in particular the Scanning Law (see the \gaia\ technical note GAIA-C3-TN-LU-LL-124-01). 
We address this potential concern in two key ways. 
First, we primarily focus on bright stars (G < 18) that are less affected by this systematic. 
Second, we structure our ultimate comparisons of data to our models (as described in later sections) around a large spatial coverage, significantly larger than the spatial scale of the systematic error fluctuations.

\subsection{Mock Data Creation}\label{ssec:modeldes}

To create the mock SMC data set, we make the assumption that the RG population can be approximately modeled using a simple Gaussian distribution for each spatial dimension. 
For the dimensions in the plane of the sky, we use the values from the earlier 2D Gaussian fit to the RG \gaia\ data. 
We find a standard deviation of approximately 1.3 kpc for the longer axis and 1.0 kpc for the shorter axis with a rotation of $\sim 55$ degrees. 
For the line-of-sight (LOS) depth, we refer to \cite{jacyszyn17} for their measurements of the RR Lyrae distances and axial ratios in the SMC. 
Additionally, as has been shown, the range of LOS distances varies in the SMC as a function of spatial position. 
To account for this, we vary the Gaussian mean of the distance distribution as a function of $X$ and $Y$, which is able to reproduce the observed tilt in the SMC (with the southeastern corner being the closest to the observer and the northwestern corner being the furthest).

\begin{figure}
\centering
 \includegraphics[width=3.35in]{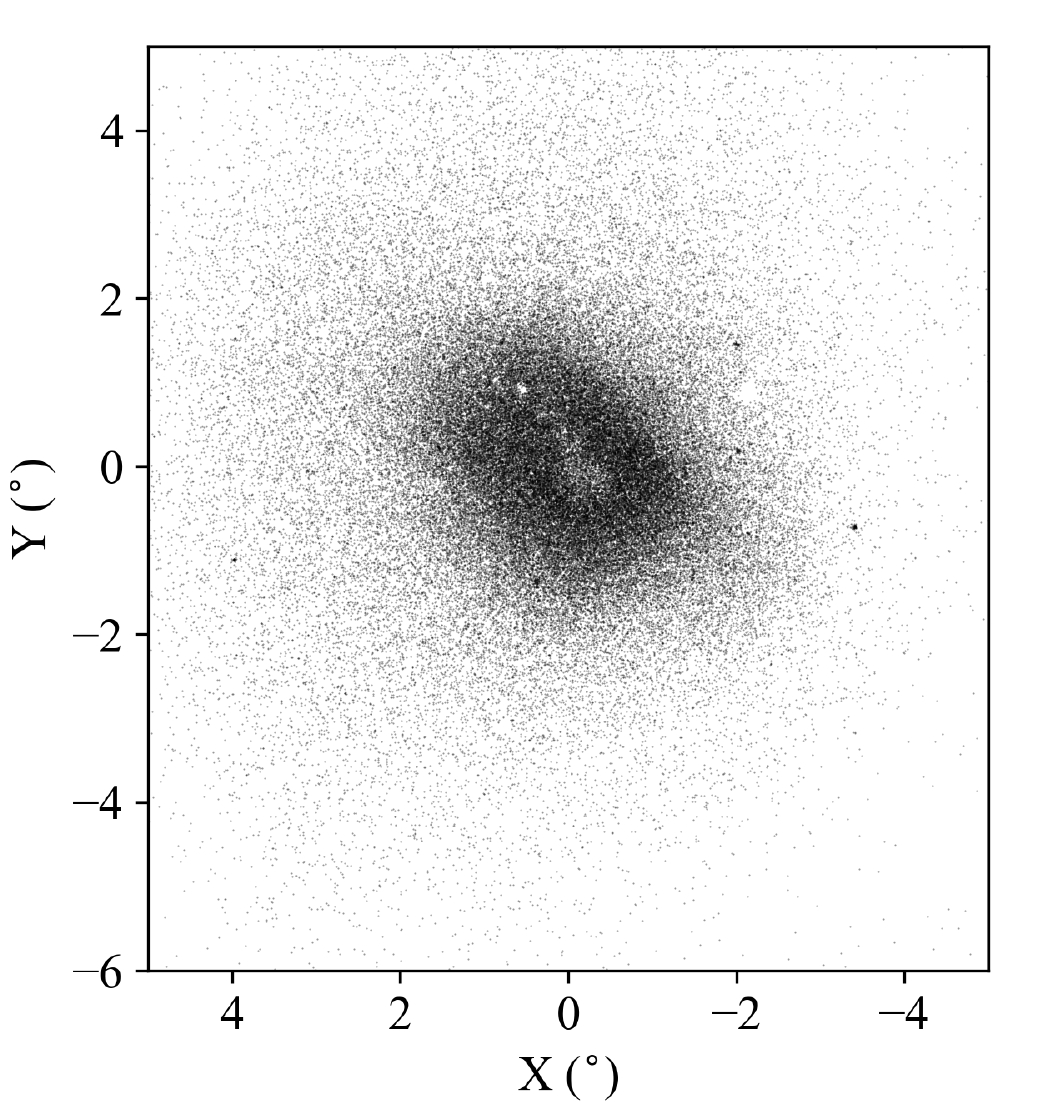} 
 \includegraphics[width=3.35in]{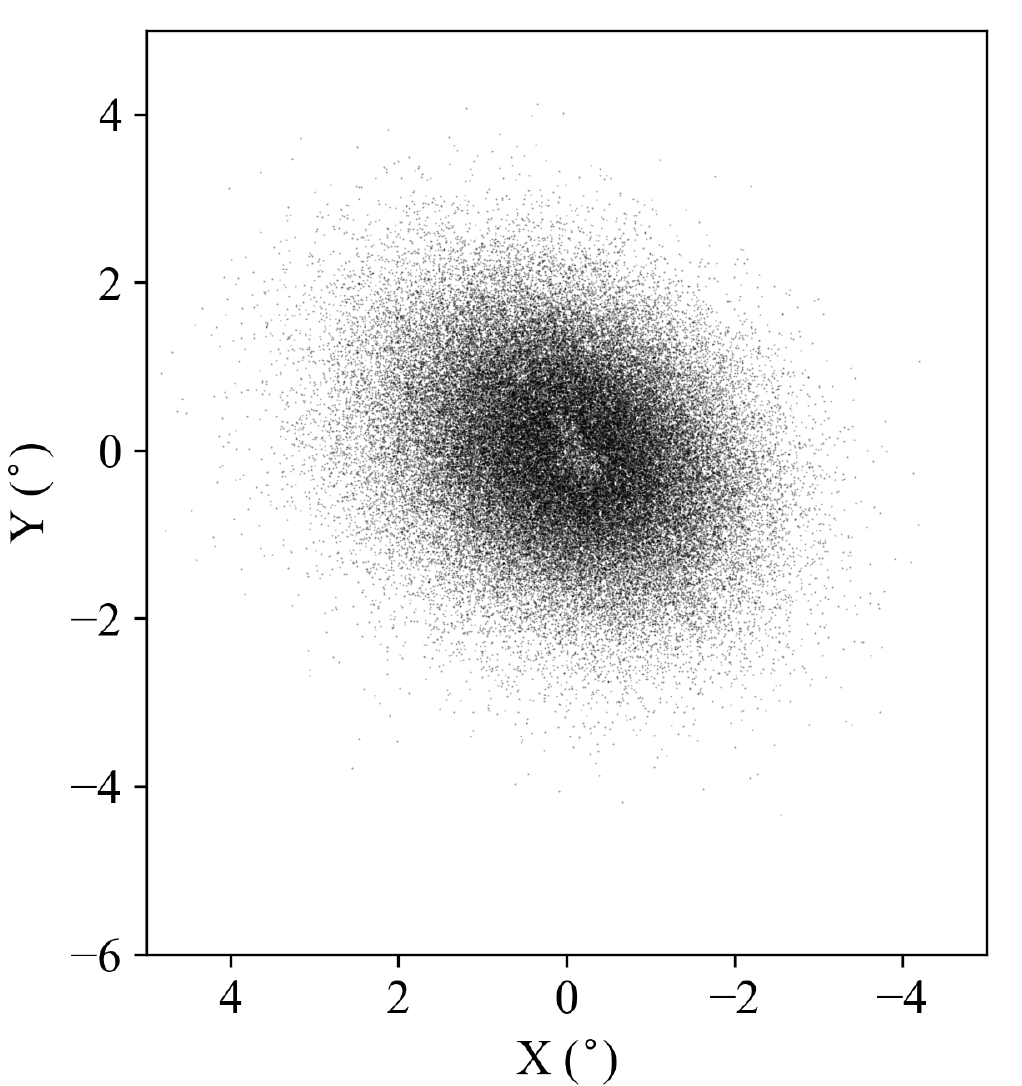} 
\caption{(\textbf{Top}) All RG stars in our selected sample plotted using our Cartesian frame. 
The small, tightly clumped dark areas in the plot are believed to be globular clusters or other stellar associations belonging to the SMC. 
(\textbf{Bottom}) All model stars, scaled to qualitatively match the observational data, plotted in our Cartesian frame. The central underdensity present in the data has been artificially created in the model in an effort to reflect observations.}
\label{f:poscomp}
\end{figure}

We then apply a random subsampling to the central region of our model to reflect the loss in astrometrically well-behaved stars observed in the RG \gaia\ data. 
At this point we note that in the \gaia\ data there appears to be two different stellar components: a dense core, relatively well-fit by a 2D Gaussian, and an offset halo of stars (offset as it appears to extend significantly further to the northeast). 
As this may represent a separate component of the SMC with unique kinematic properties, we choose to only model the dense core of the SMC, resulting in the model seen in Figure \ref{f:poscomp}. 
We also observe that our modeling does not quite capture the apparent boxy-ness of the observed data. 
However, as we are focused on developing a simple physical intuition for SMC kinematics, we believe the model to be appropriate for our purposes.

\defcitealias{vdM02}{vdM02}

With a 3D spatial model complete, we turn to assigning velocities to each of our mock stars. For this, we adopt the formalism outlined in \cite{vdM02}(hereafter \citetalias{vdM02}) to describe observations of a rotating solid body whose center of mass (COM) has some non-zero 3D motion. For more description, we refer the reader to that work. Here we will provide an outline of the key concepts used in creating the kinematic model, including the mechanisms described earlier. The kinematics for each star in our model can be described with the following:

\begin{equation}
    \begin{pmatrix}
    v_1 \\
    v_2 \\
    v_3 \\
    \end{pmatrix}
    =
    \begin{pmatrix}
    v_1 \\
    v_2 \\
    v_3 \\
    \end{pmatrix}_{\mathrm{CM}}
    +
    \begin{pmatrix}
    v_1 \\
    v_2 \\
    v_3 \\
    \end{pmatrix}_{\mathrm{int}}
    +
    \begin{pmatrix}
    v_1 \\
    v_2 \\
    v_3 \\
    \end{pmatrix}_{\mathrm{pn}}
    +
        \begin{pmatrix}
    v_1 \\
    v_2 \\
    v_3 \\
    \end{pmatrix}_{\mathrm{tidal}},
\end{equation}
where the different velocity components are: $v_{\mathrm{CM}}$, the center-of-mass motion of the SMC; $v_{\mathrm{pn}}$, the precession and nutation of the velocity plane of the SMC; $v_{\mathrm{int}}$, the internal rotation and dispersion of the SMC; and $v_{\mathrm{tidal}}$, the tidal expansion due to the interactions with the LMC.

For the initial ($x, y, z$) frame, we orient it such that ($0, 0, 0$) is located at the dynamical center of the SMC, with the positive $z$-axis pointing in the direction of the observer, the positive $x$-axis towards the west, and the positive $y$-axis pointing north.
We first assign ($x, y, z$) coordinates for each star in our SMC model then use these coordinates to calculate angular coordinates in the frame of the observer (distance, $\rho, \phi$), following Eq. 2 from \citetalias{vdM02}. 
These coordinates inform the decomposition of the different velocity Cartesian components into the ($v_1, v_2, v_3$) frame (hereafter the $v_i$ frame).

To calculate $v_{\mathrm{CM}}$, the model is created with an initial kinematic center ($\alpha_0, \delta_0$, $D_0$) and associated motion ($\mu_{\alpha}, \mu_{\delta}, v_{\mathrm{sys}}$). 
Using these quantities, the transverse velocity ($v_t$) and direction of the transverse velocity ($\Theta_t$, measured east over north) are calculated. 
Combined with the star's angular coordinates, we use Eq. 13 from \citetalias{vdM02} to calculate the velocities for each star given its angular position on the sky.

For $v_{\mathrm{int}}$, we assume a linearly increasing rotation curve, with a maximum velocity of $V_0$ at a scale radius of $R_0$, after which the rotation flattens out, as described in Equation \ref{eq:vrot}. 

\begin{equation}\label{eq:vrot}
    V(R') = \begin{cases} 
      \frac{R'}{R_0} V_0 & R' < R_0 \\
      V_0 & R' \geq R_0
   \end{cases}
\end{equation}
For simplicity, we assume cylindrical rotation, where $R'$ is only dependent on the $x'$ and $y'$ coordinates of the star in the rotating frame, denoted by $'$. 
While this likely represents an oversimplification of rotation in older, at least partially dispersion-supported, systems, we choose this prescription for consistency with other prior analyses of SMC rotation. 
For the plane of rotation, we define an inclination $i$ and a LON PA $\theta$ to describe the orientation of the plane with respect to the internal spatial frame of the galaxy. 
We then convert the internal Cartesian vectors to the $v_i$ frame using Eq. 21 from \citetalias{vdM02}. 
Once in the $v_i$ frame, we add a velocity ``kick'' to each component, randomly drawn from a normal distribution using 26 km s$^{-1}$ (from \citealt{dobbie14a}) as a standard deviation. 


As a brief aside related to the internal motions, we note that we are aiming to only model the observable motions of the SMC. 
As such, we are not considering asymmetric drift when reporting our final rotation curves. 
For the SMC, the known dispersion in the RG population of $>20$ km s$^{-1}$ would require significant corrections in order to model the rotation curve predicted by the underlying gravitational potential.
However, the observational impact of asymmetric drift on older stellar populations, the flattening of the rotation curve beyond a certain scale radius due to random kicks and perturbations experienced by individual stars, mirrors the model we apply for rotation. 
As such, our modeled rotation can be understood as the curve measured for non-drift-corrected motion, just as one would measure in the raw \gaia\ data as well.


For $v_{\mathrm{pn}}$, knowledge of the time dependency of $d\theta/dt$ and $di/dt$ is required. 
For the SMC, there are no known constraints for $d\theta/dt$, so we choose to not include it in the modeling. 
Attempts have been made at constraining $di/dt$ in the SMC \citep[e.g.,][]{diteodoro19, dobbie14a}, but the parameter space is still fairly unconstrained. 
Additionally, for an object the size of the SMC, contributions from even a fairly large $di/dt$ would only account for up to a few km s$^{-1}$ in the radial velocity component, significantly smaller than our uncertainties from other factors. 
As such we choose to not include $v_{\mathrm{pn}}$ in our final calculation.

For $v_{\mathrm{tidal}}$, given the complex gravitational interactions between the SMC and LMC, we do not attempt a rigorous numerical treatment. 
Instead, we use the expectation that tidal expansion should occur along the direction of relative motion between the two bodies to create a prescription for the individual stars, as seen in numerical simulations of dispersion-supported systems on plunging orbits into larger hosts \citep[e.g.,][]{penarrubia09}.
In this scenario, stars are evacuated out of the satellite across all locations, eventually forming wide tidal tails on either side of the satellite, rather than only through the Lagrange points into thin, kinematically cold tails, as observed in tidally disrupted globular clusters in the MW. 
As described earlier in Section \ref{ssec:obsdata}, we calculate the relative velocity, $v_{\mathrm{rel}}$ between the LMC and SMC to determine the direction of the tidal component, which for simplifying purposes we will assume to be fixed for every star.
For the magnitude of the motion for an individual star, we use the distance from the SMC center along $v_{\mathrm{rel}}$, $d_{\mathrm{rel}}$,
in conjunction with a scaling ratio between of 0 and 15 km s$^{-1}$ kpc$^{-1}$, $v_{\mathrm{scale}}$ (the possible range based on the relative motions in the Magellanic Bridge in \citealt{zivick19}), giving the final tidal contribution for each star, displayed below in Equation \ref{eq:tidal}.

\begin{equation}\label{eq:tidal}
    \vec{v}_{tidal} = d_{\mathrm{rel}} \cdot v_{\mathrm{scale}} \cdot \frac{\vec{v}_{\mathrm{rel}}}{\left|\left|\vec{v}_{\mathrm{rel}}\right|\right|}
\end{equation}

With all of the components combined together in the $v_i$ frame, we then convert these physical velocities (km s$^{-1}$) into observed motions (mas yr$^{-1}$ for $\mu_{\alpha}$ and $\mu_{\delta}$) using Eq. 9 from \citetalias{vdM02}. 
We add a random kick, selected from a Gaussian distribution with standard deviation of 0.15 mas yr$^{-1}$, calculated using the RG PM errors in Section \ref{sec:data}, to the measured $\mu_{\alpha}$ and $\mu_{\delta}$, to reflect the uncertainties in the \gaia\ DR2 catalog. 
This results in a final catalog for our model of stars, with each one possessing (RA, Dec, $\mu_{\alpha}$, $\mu_{\delta}$, $v_{\mathrm{LoS}}$) comparable to the data.

\subsection{Comparison Quantification}\label{ssec:modelquant}

To quantify our data, we require that our model be able to reproduce both the magnitude of the residual motions and the location of the residual motions. 
To that end, we construct a set of spatial versus kinematic comparisons.
As we have chosen to only focus on the \gaia\ PMs, ignoring the RVs due to the uncertainties in the underlying distance distribution of the sample, and as we do not have distance information for individual stars, we are limited to four comparisons: $\mu_{x}$ vs $x$, $\mu_{y}$ vs $x$, $\mu_{x}$ vs $y$, and $\mu_{y}$ vs $y$.

For a choice of the $x/y$ frame, we adopt the SMC's shape as a natural frame, assigning the $x$-axis to align with the major axis and the $y$-axis to align with the minor axis of the SMC RG distribution (as seen by the dashed black lines in Figure \ref{f:respm_data}). 
After rotating the positions and PMs of the RGs into this new frame, we bin the PMs along the spatial axis, using the Freedman-Draconis (FD) Rule to set the bin width \citep{FD81} to avoid over-fitting the data. 
We set the bin widths using the observational data but limit the comparison between model and data to only bins containing model data.

\begin{figure*}
\centering
 \includegraphics[width=3.3in]{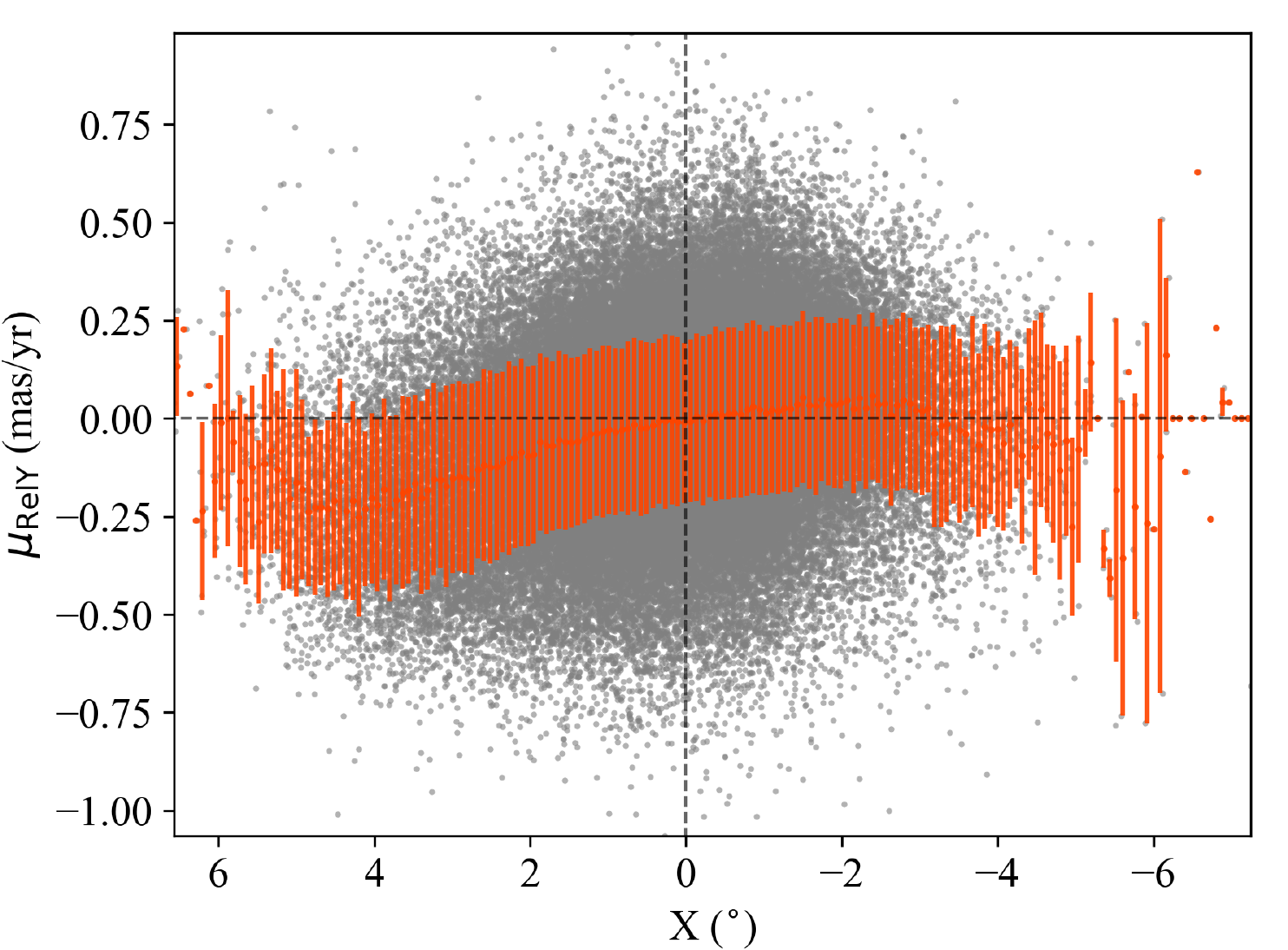} 
 \includegraphics[width=3.3in]{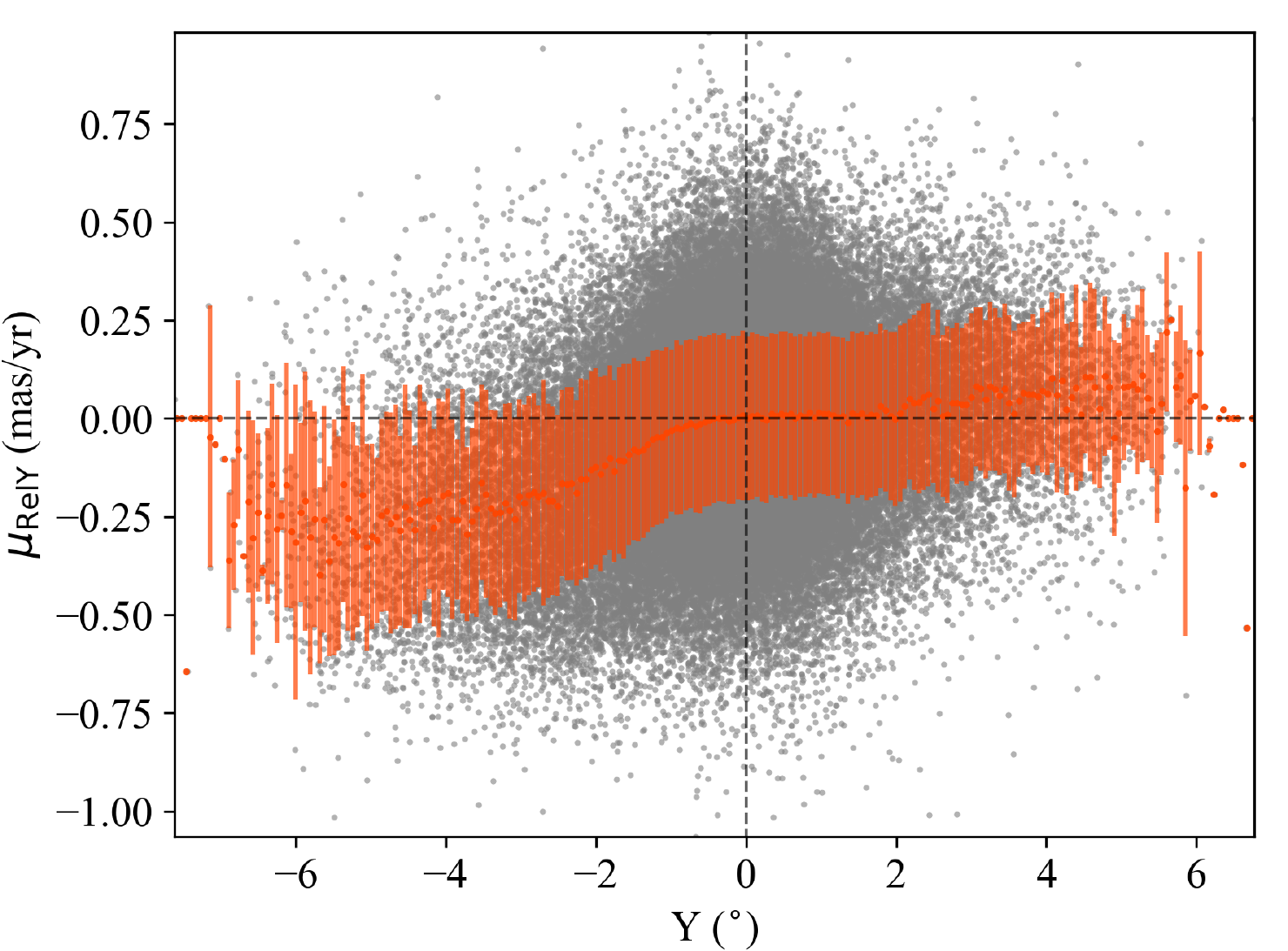} 
 \caption{Residual PMs plotted against spatial position in the SMC geometric major/minor axis frame. The gray points are all of the RG stars in the same. The stars are binned accordingly to the spatial sampling, which is about 0.08 degrees in X and 0.06 degrees in Y. A simple average is taken for each bin, marked by the orange-red point, and the standard deviation of the bin is shown as the error bar. \textbf{(Left)} Residual PM in Y (which points roughly along the axis of relative LMC-SMC motion) as a function of position in X (which spans the major axis of the SMC, with the positive X direction north of the Magellanic Bridge). \textbf{(Right)} Residual PM in Y as a function of position in Y.}
   \label{f:data_spatkin}
\end{figure*}

For each bin, we calculate the average of the residual PMs and the variance in the bin. 
Two examples of this spatial-kinematic scheme can be seen in Figure \ref{f:data_spatkin}, which displays only the \gaia\ RG data.
Then we calculate a $\chi^2$ value between the mock data from each proposed kinematic model and the \gaia\ RG data for each of the four spatial-kinematic permutations.
Bins with fewer than 5 stars are excluded from this calculation to avoid unintentionally biasing the comparisons towards sparsely sampled outer regions of the SMC model. 
To compare between different models, the four separate $\chi^2$ values for a given model are summed to produce a single statistic that is then minimized.
Figure \ref{f:modelcomp} demonstrates this comparison, with the mock data from an example model overplotted on the underlying data (each with binned averages and the standard error of the mean), with the spatial range being narrower than the data-only spatial-kinematic distributions from Figure \ref{f:data_spatkin} (due to our focus on the SMC core and not the extended halo structure).

\section{Model-to-Data Comparison Results}\label{sec:comp}

In exploring the parameter space, 
we choose to focus our analysis on four key parameters of the model: inclination angle $i$, the position angle of the Line of Nodes (LON) $\Theta$, the rotation velocity $V_0$, 
and the tidal expansion scale factor $v_{\mathrm{scale}}$. 
Since we are utilizing a simple 3D model for the SMC and a forward modeling approach aimed at building physical intuition for the SMC kinematics, we simply test the parameter space in discrete steps for each of the parameters.

For the remainder of the parameters, we choose to keep the distance $D_0$, the proper motion ($\mu_W$, $\mu_N$), center ($\alpha_0$, $\delta_0$), the systemic velocity $V_{sys}$, and the rotation scale radius $R_0$ fixed.
For the distance and systemic velocity, both have already been better established for the RG population in prior studies \cite[e.g.,]{jacyszyn17, dobbie14a}. 
For the center and systemic PM, as we have already measured them based on the data, we choose not to test them to avoid introducing unexpected biases in our $\chi^2$ calculation.
For the rotation scale radius, we choose to approximate it by using the radius at the initial peak of $\mu_{\theta}$ versus radius in Figure \ref{f:r_vt_data}, at roughly $\sim1$ degree (which corresponds to $\sim1$ kpc at a distance of 60 kpc).
Given the relatively flat behavior of $\mu_{\theta}$ (with small perturbations) for much of the inner radius until the large increase in the outer region, consistent with the effects of tidal expansion dominating, we think this scale radius is appropriate for our analysis.
Finally, as described earlier, given the large degree of uncertainty and relatively small effect of the precession/nutation, $di/dt$, we do not include it.

The parameter space step gradation for each parameter is as follows:
\begin{itemize}
    \item $i$: every 10 degrees from 0 to 90 degrees (rotation in plane of sky vs edge-on rotation)
    \item $\Theta$: every 10 degrees from 0 to 360 degrees
    \item $V_0$: every 5 km s$^{-1}$ from 0 to 30 km s$^{-1}$, which probes the lower to upper range of previous inferred values for the SMC stellar component
    \item $v_{\mathrm{scale}}$: every 5 km s$^{-1}$ kpc$^{-1}$ from 0 km s$^{-1}$ kpc$^{-1}$ to 15 km s$^{-1}$ kpc$^{-1}$
\end{itemize}

After creating a mock set of data for each permutation of the above parameter space, the summed $\chi^2$ values were compared.
As this was not a continuous exploration of the parameter space but a discrete one, we aim to identify a class of best-fitting models and examine the commonalities among them to build our understanding of the SMC.
For this, we bin the $\chi^2$ values using the FD rule and calculate the gradient across the bins, choosing the largest positive gradient as the cutoff for the best-fitting class of models.
Using this criterion results in roughly 430 models selected out of the over 12,000 tested.

\begin{figure*}
\centering
\begin{tabular}{cc}
    \includegraphics[width=3.3in]{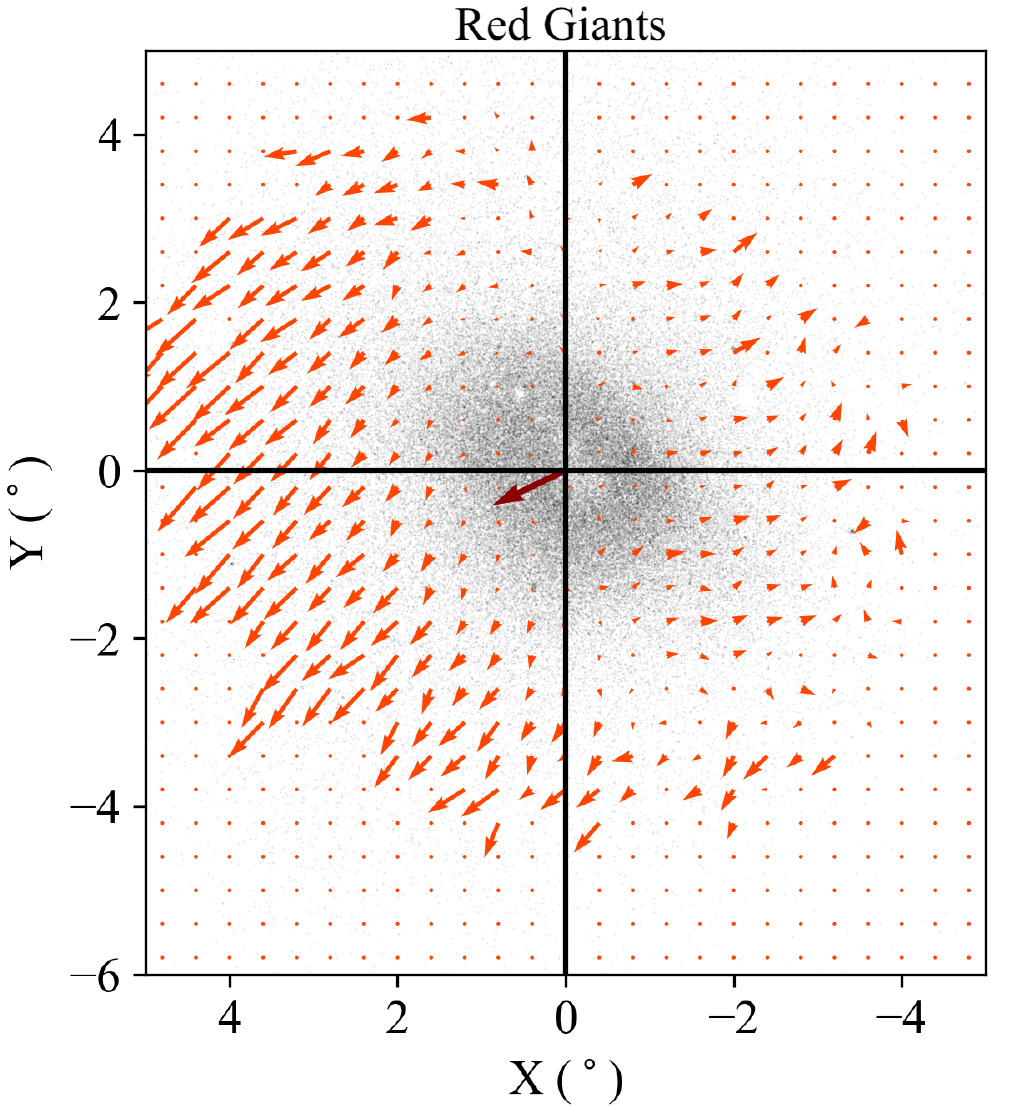} & 
    \includegraphics[width=3.3in]{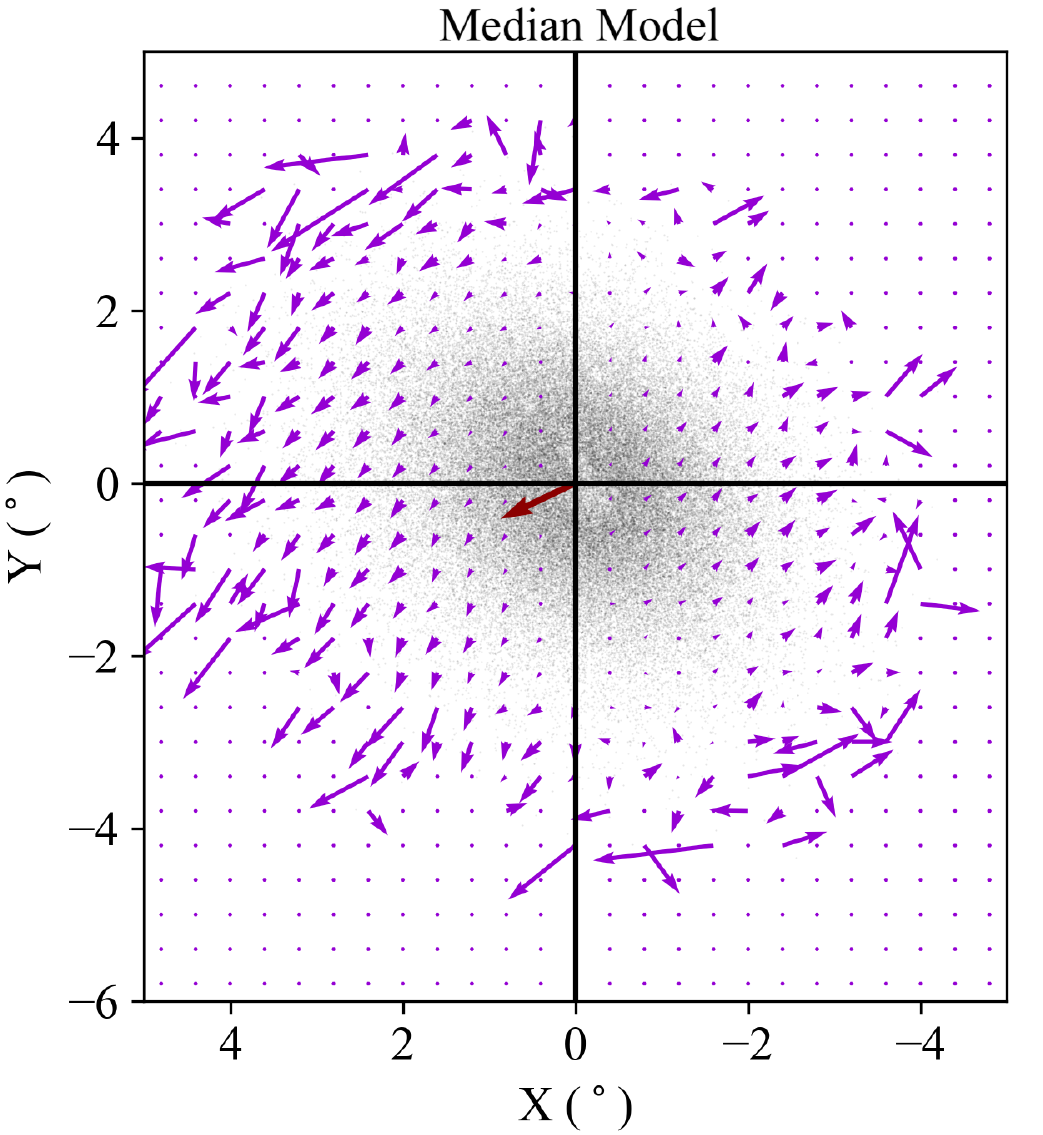} \\
\end{tabular}
\caption{Residual PM vector plots for the \gaia\ RG stars (orange-red, left) and the best-fit model (purple, right). The frame and vectors are displayed in the same manner as Figure \ref{f:respm_data}. The \gaia\ -measured residual PMs have been limited to only the bins that have measured PMs in the mock SMC data, due to the model only including a single 2D Gaussian function that does not extend fully into the halo of the SMC. We note that the noisier signal in the model at $\sim3$ degrees from the center, compared to the \gaia\ data, results from a sparsity of stars (due to the model being constructed only to match the central SMC region) rather than an underlying physical mechanism.
         \label{f:bestfit}
         }
\end{figure*}

Across this class of best-fitting models, we find a need for both a non-zero rotation ($V_0$ between 15-25 km s$^{-1}$, with the assumed $R_0$ of 1 kpc, placing the rotation likely below the measured dispersion of the SMC RGs but notably non-zero), at a relatively high inclination ($i$ between 70-90 degrees) with a LON oriented from north to south ($\Theta$ around 150-230 degrees), and a non-zero tidal expansion component ($v_{\mathrm{scale}}$ near 10 km s$^{-1}$ kpc$^{-1}$).
The median parameter values from the best-fit class and comparisons to literature values can be found in Table \ref{tab:modelparam}. Estimates on the uncertainty in the model fitting can be found in Table \ref{tab:modelerror}, where the errors have been estimated using the median absolute deviation (MAD).
Figure \ref{f:bestfit} shows the residual vector plot for the median parameters of the best-fit class of models, which displays similar key characteristics to the \gaia\ data: large and roughly linear motion in the eastern portion of the SMC and a weak signal of rotation within the inner 1 degree of the spatial distribution.

\begin{figure*}
\centering
\begin{tabular}{cc}
    \includegraphics[width=3.3in]{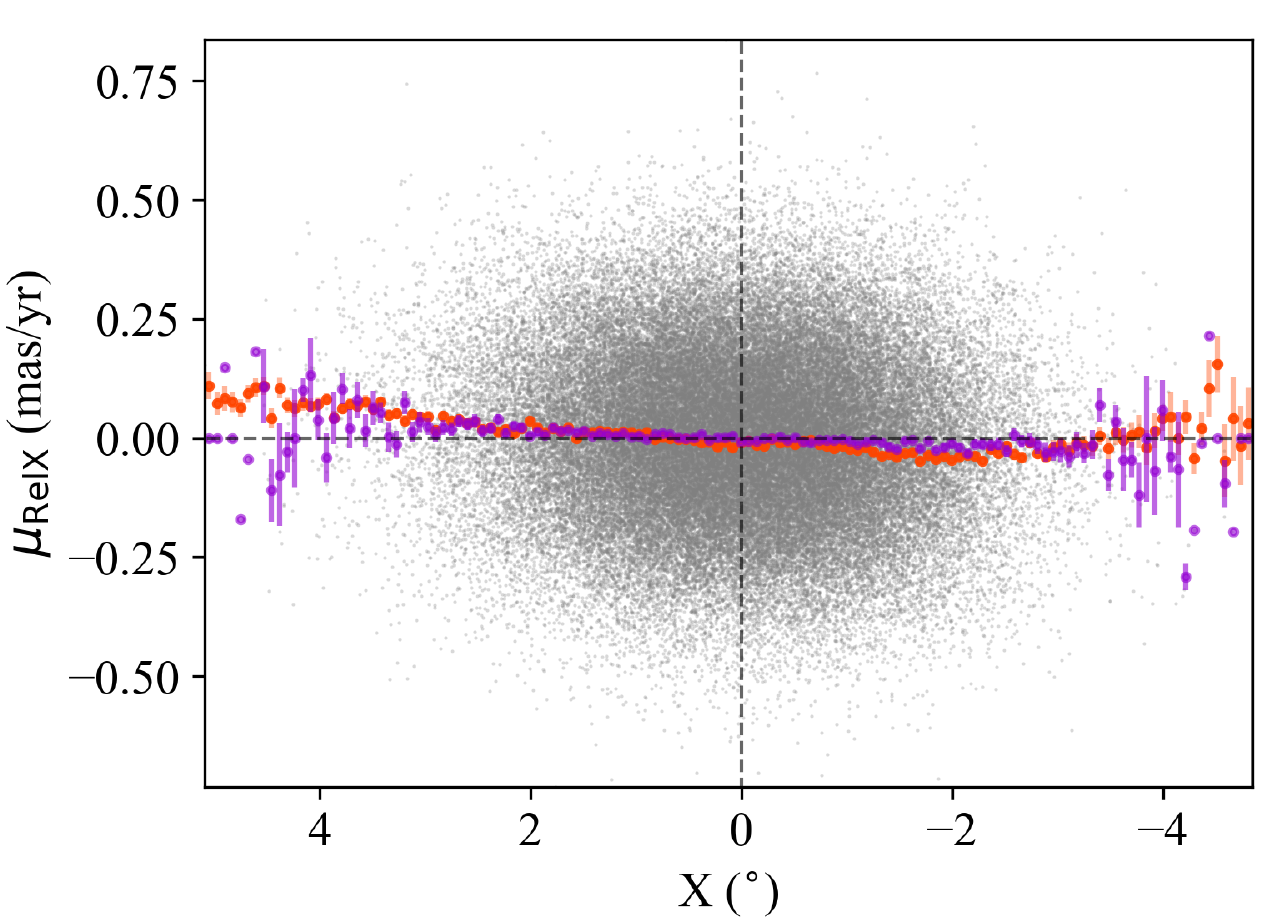} & 
    \includegraphics[width=3.3in]{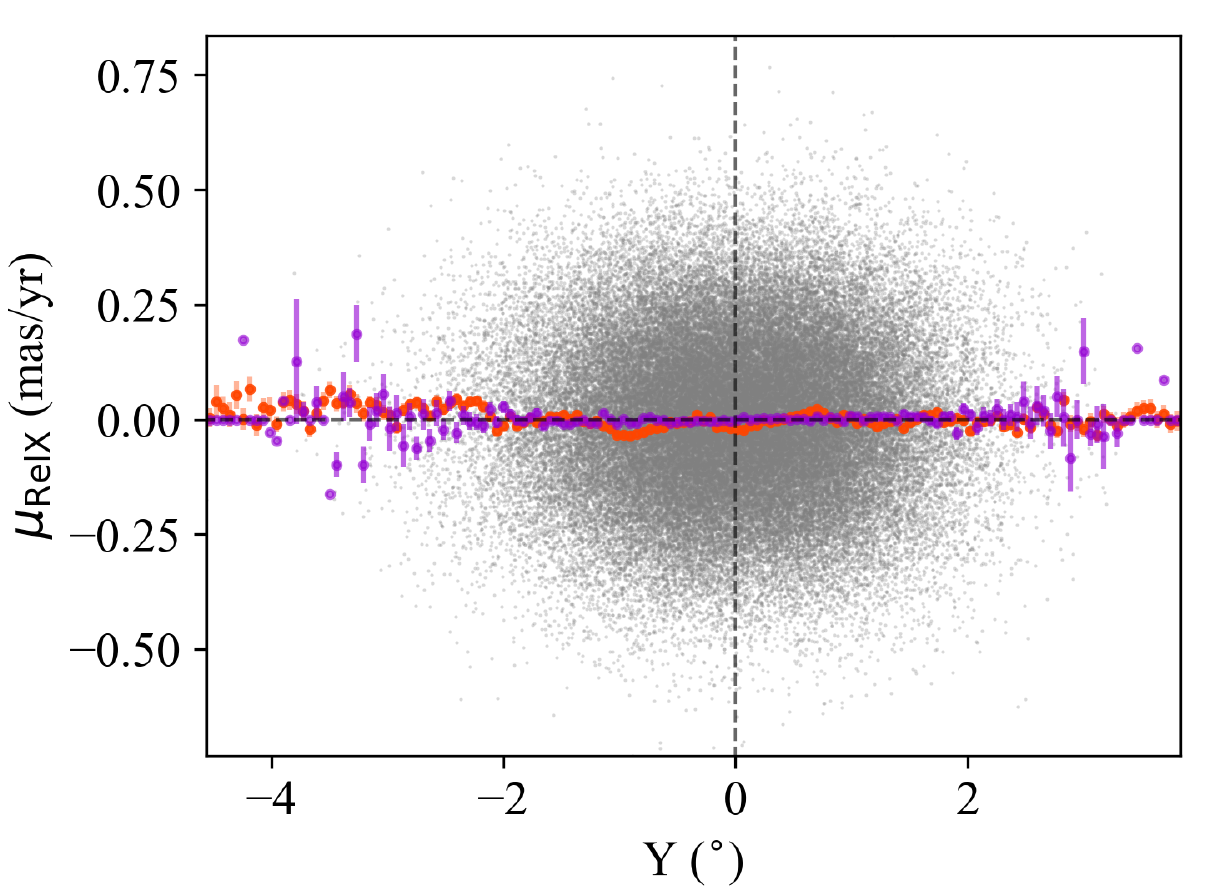} \\
    \includegraphics[width=3.3in]{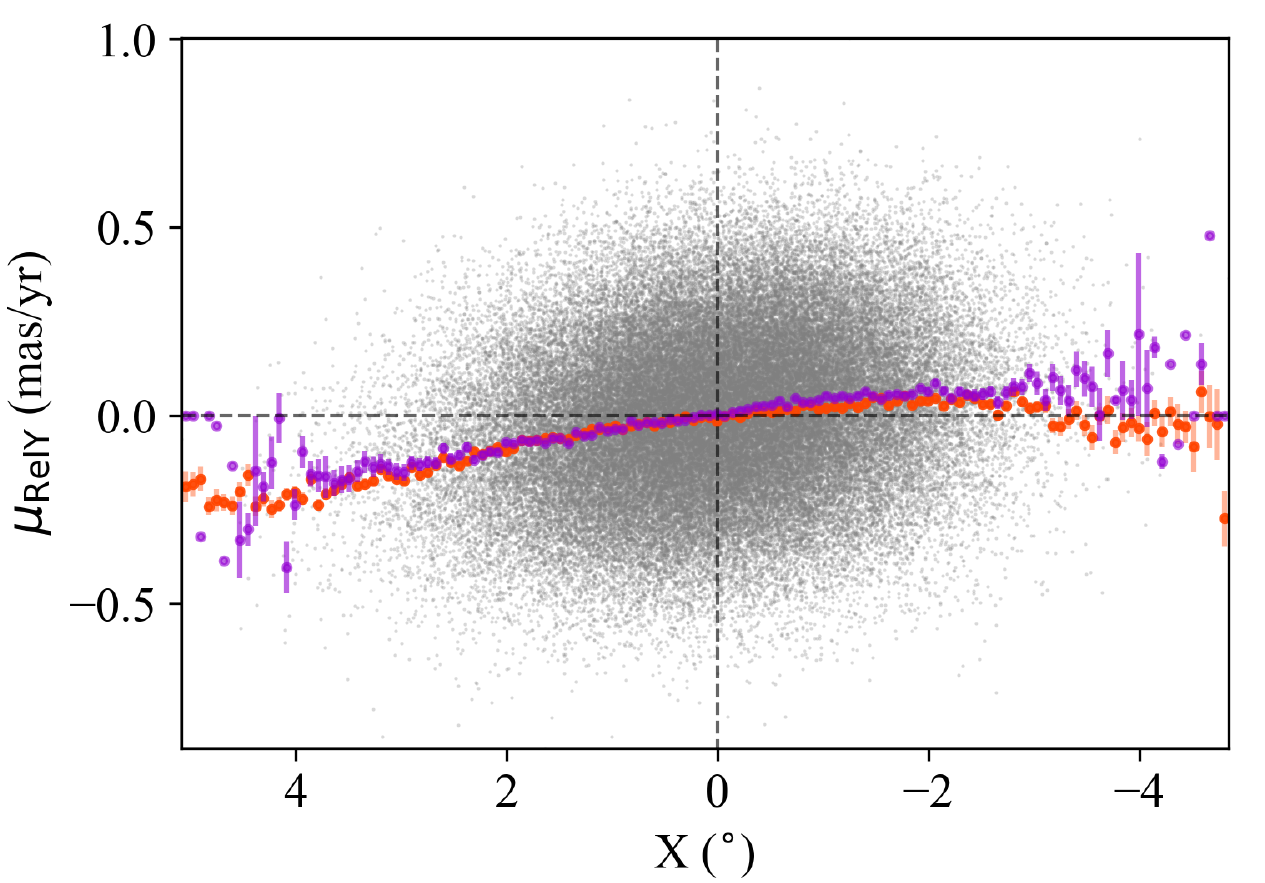} &
    \includegraphics[width=3.3in]{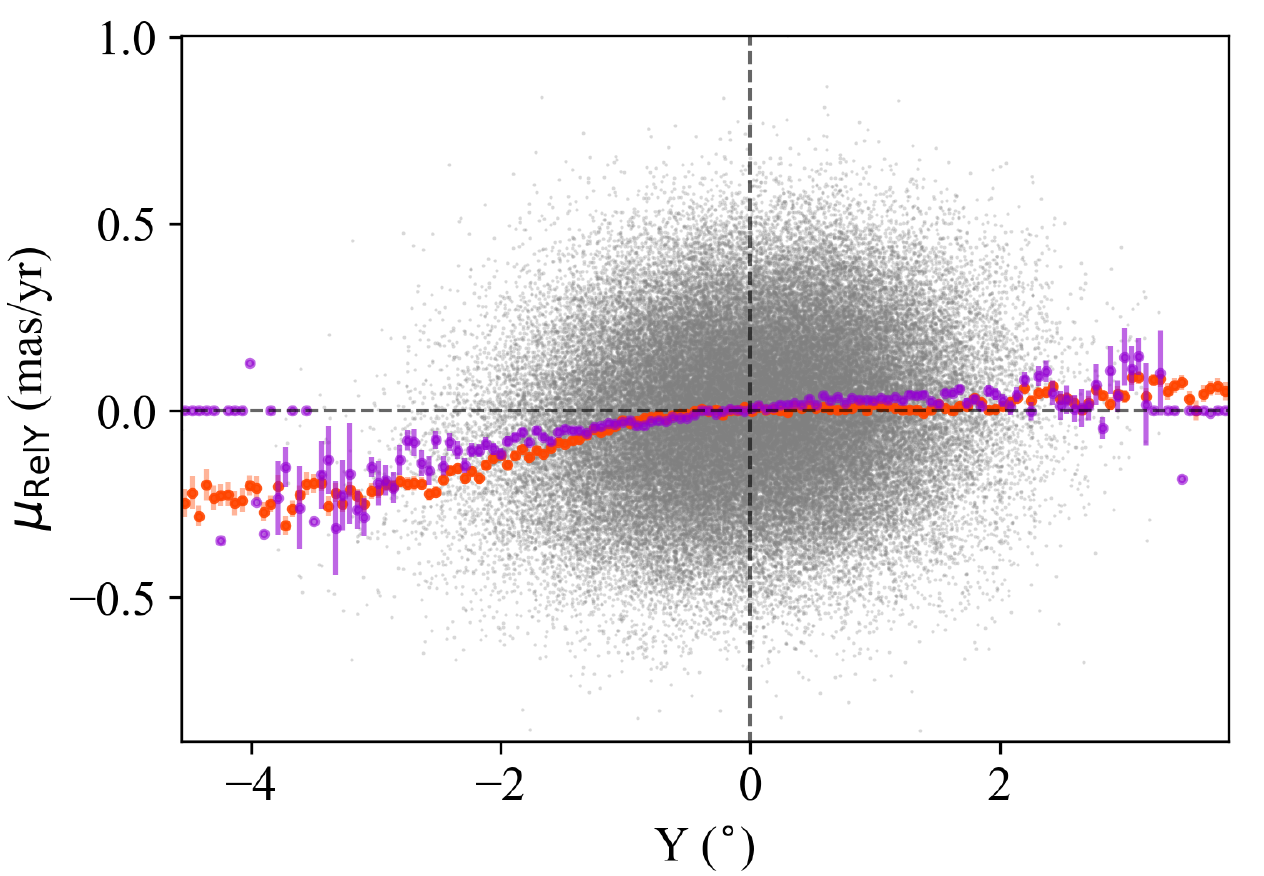}
\end{tabular}
\caption{Residual PMs as a function of spatial position for all permutations of PM$_X$/PM$_Y$ and $X/Y$. The light grey points are the mock RG stars from the kinematic models. The orange-red points are the measured \gaia\ RG residual PM averages, and the purple points are the averaged residual PMs for the mock RG stars. The model presented here is the best-fit model to the data (specific parameters are listed in Table \ref{tab:modelparam}). 
         \label{f:modelcomp}
         }
\end{figure*}

We can see this in the spatial-kinematic comparisons in Figure \ref{f:modelcomp}, where our model is able to capture the behavior across much of the SMC, including asymmetric behavior in the $y$ vs $\mu_Y$ space. 
This asymmetry similarity can be seen in the residual vector plots as well (Figure \ref{f:bestfit}) where the vectors on the eastern side display significantly larger magnitudes than the vectors on the western side. 
We primarily attribute this behavior to a combination of a two different factors.

First, our distance-dependent prescription for tidal expansion combined with the relative LMC-SMC vector and the 3D structure of the SMC lead to a maximizing of the tidal contribution.
This is due to the southeastern-northwestern axis having the largest LOS depth, which coincides with the direction of the relative motion, leading to larger tidal contributions than if the two directions were perpendicular (rather than roughly parallel) to each other.
That they have this alignment may speak to the evolution of the SMC and its interactions with the LMC, but more detailed analyses and comparisons to simulation work of the Clouds will be required to be able to properly evaluate the significance of this alignment.

Second, this large distance gradient in the LOS depth results in the observed PMs (regardless of underlying physical mechanisms) between the two sides being significantly different ($\sim15\%$ for stars sitting $\pm5$ kpc from the center of the SMC, $\sim30\%$ at $\pm10$ kpc).
Given the magnitude of the underlying velocities, these fractional differences can result in significant differences 
in their projection into our observed space.

We do note some deviations from the observations, particularly on the western side of the SMC.
Tied into the above discussion of the asymmetry, our model predicts slightly larger motions than observed around 2 degrees from the center of the SMC.
We also note that there appears to be a region in the southwest that has predominantly western residual PMs, in contrast to the same region in our model that predicts a northwestern motion.
In both instances, one potential explanation could be a foreground feature that would lead to a different average distance (and thus a different tidal contribution).

However, on the whole we do appear to capture the general behavior of the SMC, especially the eastern portion that appears to be well-described by our model, capturing the rotation signal layered on top of the large tidal motions, which appears as the large vectors slowly changing in direction as a function of azimuthal position.
As a final note on the comparisons, we note that because our model for the SMC only aimed to capture the central spatial density of the SMC and not the more diffuse ``halo'' of stars (easiest seen in Figure \ref{f:poscomp}), the noise in our model gets significantly worse near the edges where it can be only one or two stars providing the signal, unlike in the \gaia\ RG data where there are many more stars per bin at $\sim3$ degrees from the SMC center.

\begin{figure*}
\centering
\begin{tabular}{ccc}
    \multirow{2}{*}[0.65in]{\includegraphics[width=2.2in]{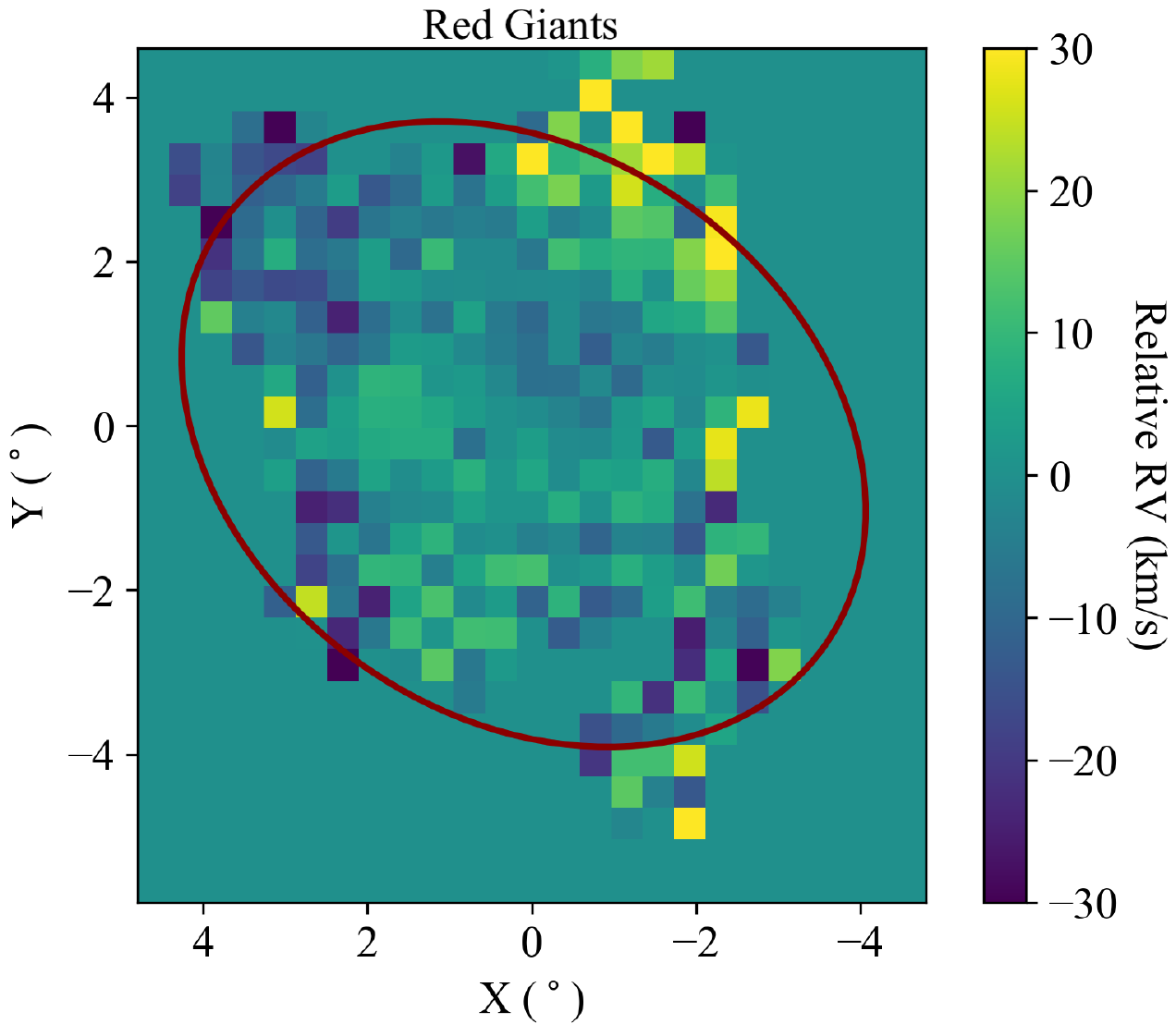}} & 
    \includegraphics[width=2.2in]{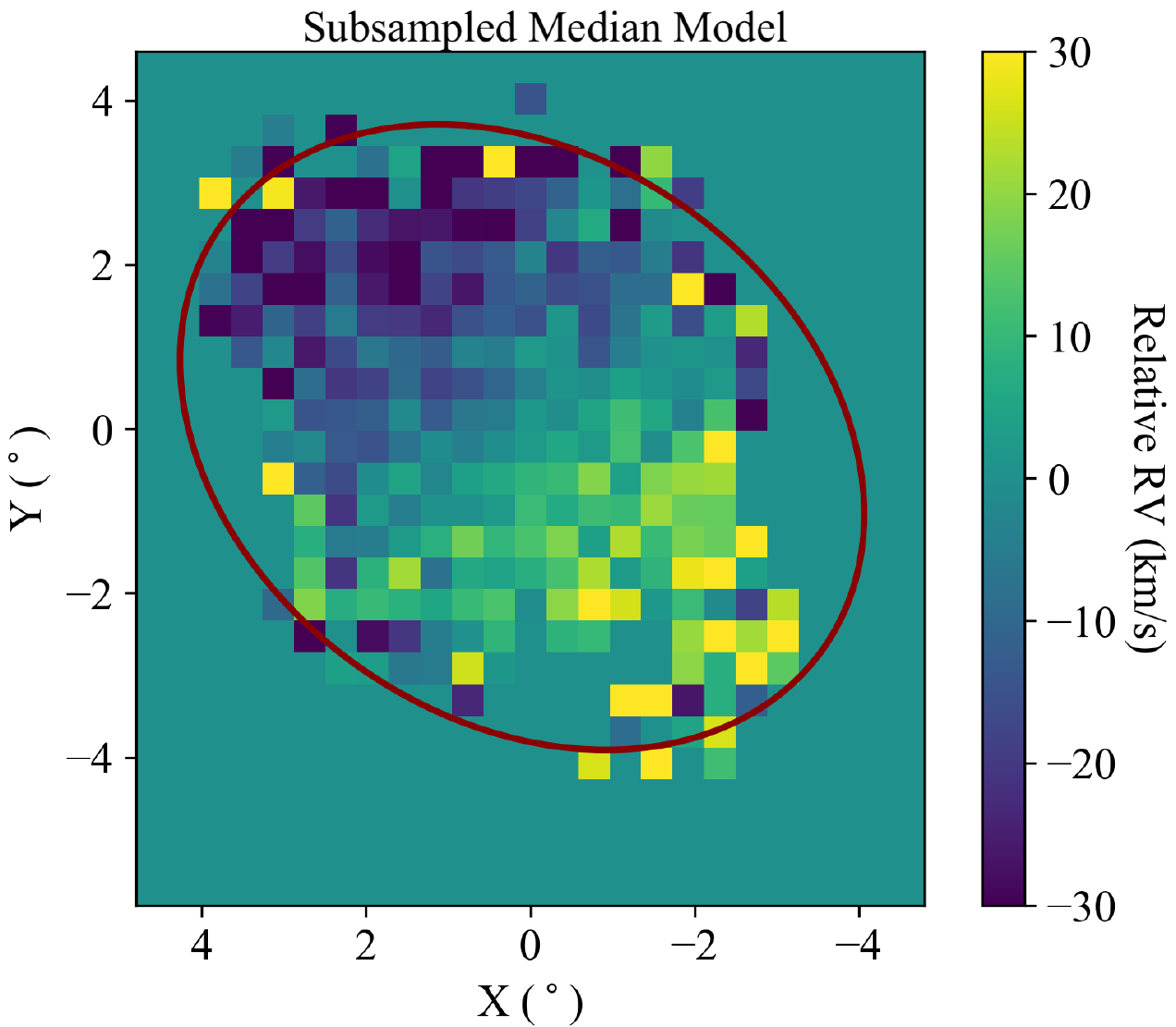} & 
    \includegraphics[width=2.2in]{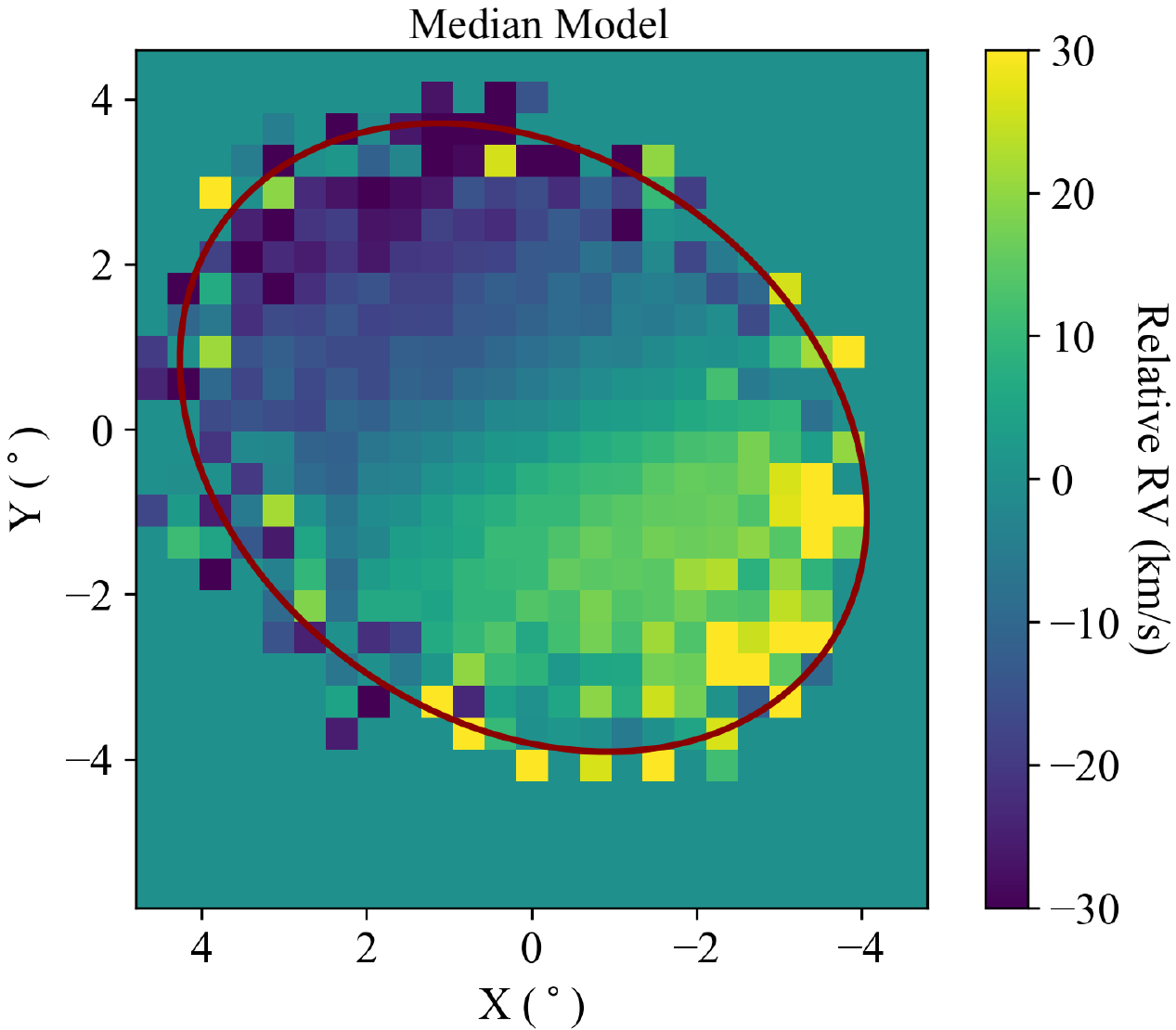} \\
    &
    \includegraphics[width=2.2in]{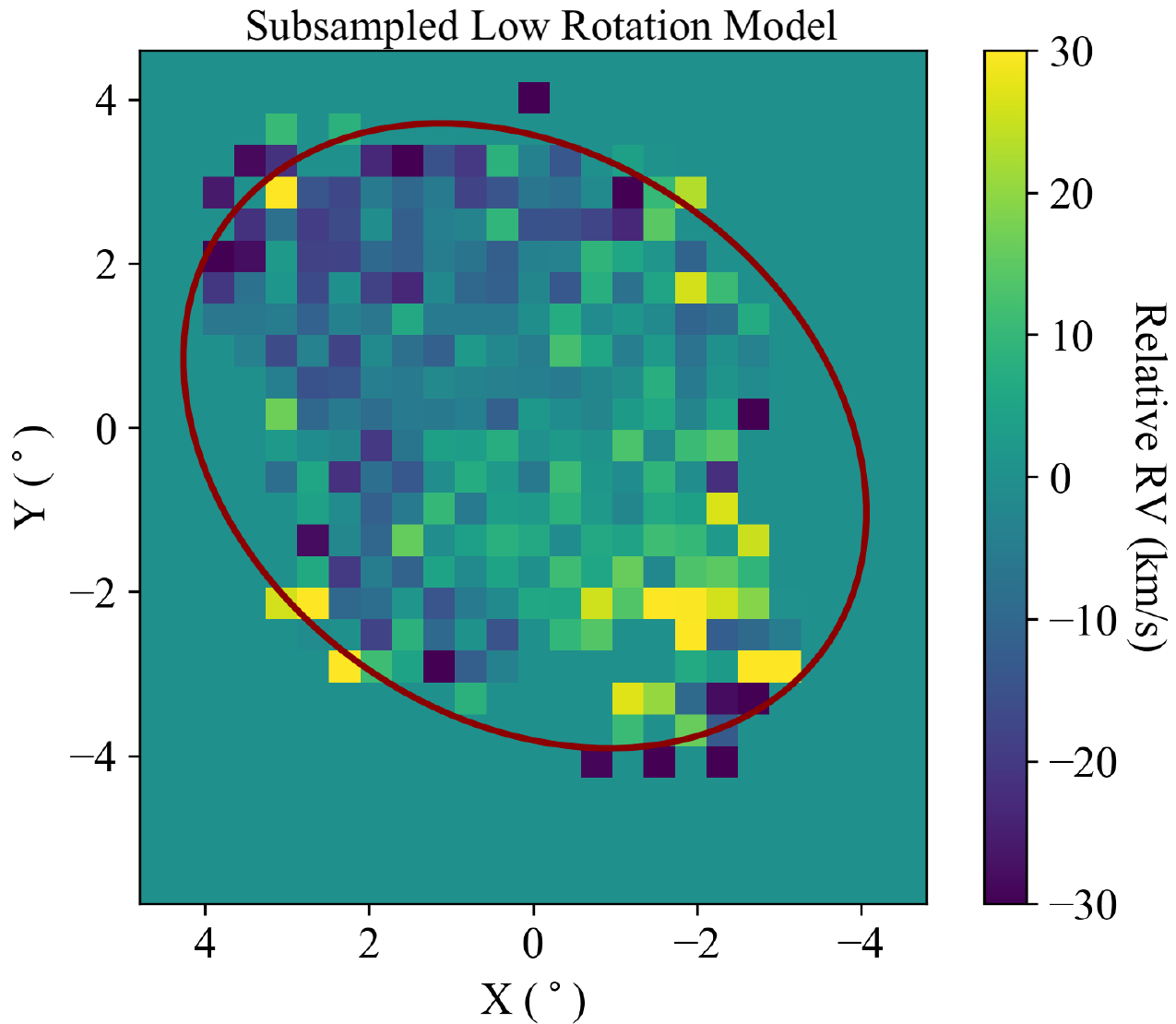} &
    \includegraphics[width=2.2in]{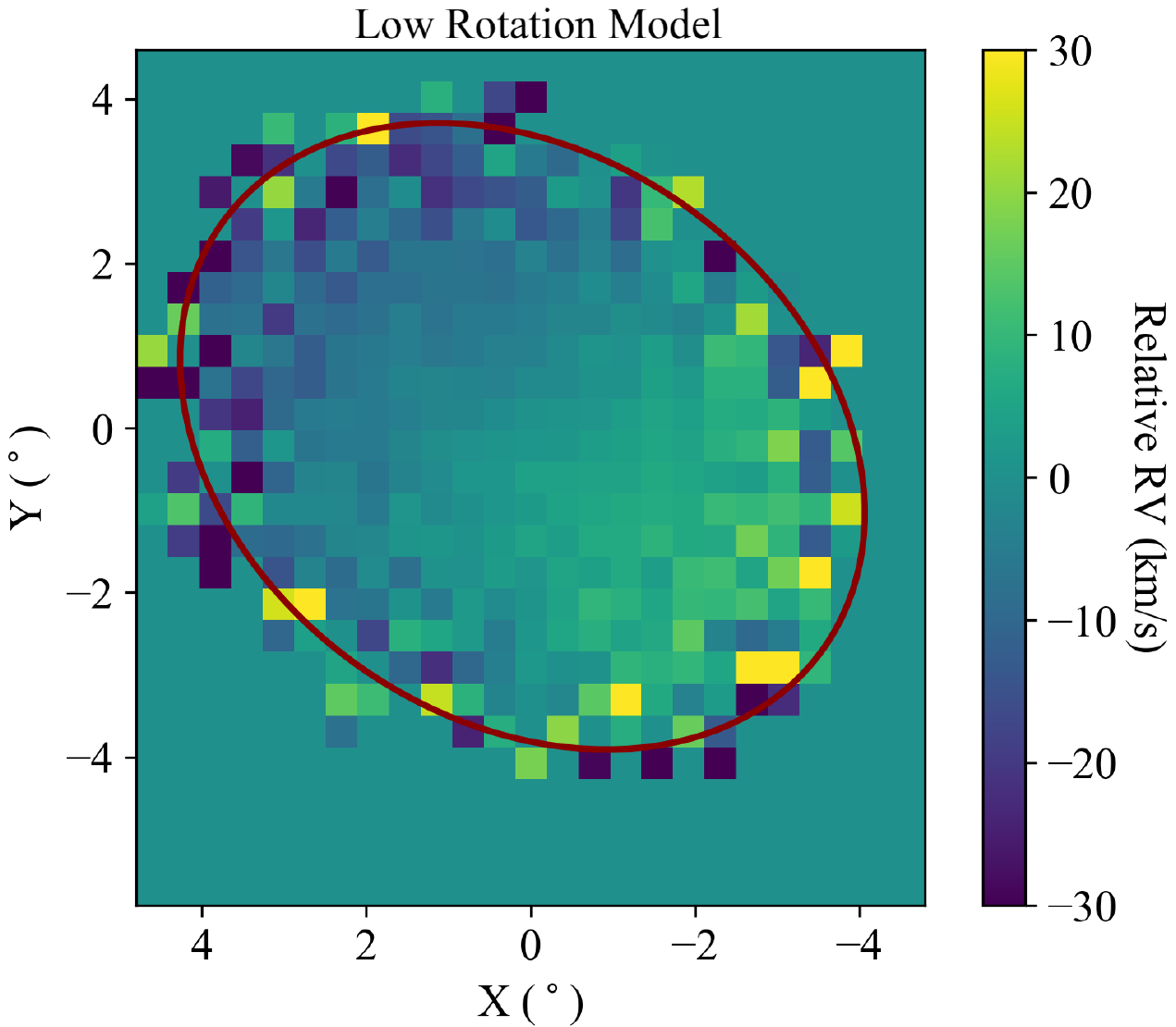} \\
\end{tabular}
\caption{\textbf{(Left column)} Binned and averaged residual radial velocities (RVs) for publicly available data , \textbf{(middle column)} our model-generated mock data subsampled to the same spatial density as the public data, \textbf{(right column)} and our full model-generated mock data. The model corresponding to the median values of the best-fit class of models is displayed in the top row of the middle and right columns, and the model with the smallest $V_0$sin$i$ from the best-fit class is displayed on the bottom row of the middle and right columns. An ellipse fit to the edges of the full mock data has been overlaid on the plots to help guide the eye in comparing spatial locations between the plots. Gaps in the spatial coverage of the available RVs can be seen, and the impact of the distance-dependence in the tidal expansion prescription can be observed in the apparent outlier values at the edges of the mock data (where stars may lie significantly in front or behind the assumed SMC center).
         \label{f:rvcomp}
         }
\end{figure*}

As a final check on our best-fit class of models, we compare our predicted RVs to the cross-matched RV catalog. 
Given the degeneracy between $V_0$ and $i$ (where larger $V_0$ can be consistent with PM constraints given a large enough $i$), for a more complete comparison to the RV catalog, we also display the model from the best-fit class with the lowest $V_0$sin$i$ as an example of predicted RVs for a slower rotating, more face-on rotation plane (the full values for this model can also be found in Table \ref{tab:modelparam}).
In Figure \ref{f:rvcomp}, we display the bin-averaged RVs for both the cross-matched \gaia\ stars (left column), mock data that has been subsampled to match the spatial density of the publicly available data (middle column with top row being the median model and the bottom the low $V_0$sin$i$ model), and the full mock data predictions (right column with same rows as the middle column).
To guide the eye in comparing directly between the plots, an ellipse has been roughly fit to the outer edges of the full mock data and overlaid on all plots.
The maximum and minimum value for the color bar have been set to emphasize the gradient within the predicted data.

When we compare the RG data to our subsampled models, we find that the slower-rotating, lower inclination model appears to be in better agreement with the data.
Qualitatively, it matches the slight gradient stretching from the northeast region of the SMC down to the south-central portion of the SMC. 
The faster-rotating, higher inclination model clearly predicts a clear gradient across the minor axis of the SMC, even when subsampled down to just 4,000 stars, which does not appear in the available RV data.
Intriguingly, when examining these two models in PM-space, the slow-rotating, low inclination model fails to reproduce the fast-rising $\mu_{\theta}$ as a function of radius that we see in the data (in Figure \ref{f:r_vt_data}), placing some amount of tension on results from PMs and results from RVs.

As one interesting note in the construction of the model, the apparent LON PA in the slower-rotating, lower inclination model would stretch from the northeast to the southwest. 
However, this mock data was generated using a LON PA of $\Theta=170$ degrees, which would normally appear from north-northwest to south-southwest.
The reason for this apparent shift lies in the underlying distance distribution of the SMC and the interplay between that distribution and our tidal prescription, where stars on the eastern side of the SMC are likely to be closer to the observer and receive a tidal contribution in the negative relative RV direction (which can be seen in the actual RV data on the central-eastern edge of the SMC).
Similarly, stars in our model in the southwest region tend to lie behind the assumed SMC center and receive a tidal contribution in the opposite direction, enhancing the expected rotation component there.

This leads to the first of two noticeable areas of deviation between the models and the data.
In the RV data, the southwest region appears to have stars at both extremes of the relative RVs there.
With this tension, it is possible that our underlying 3D model of the SMC does not accurately capture the distance distribution of the RGs here, with more stars sitting either near the SMC center or in front of the SMC center, leading to less tidal tidal contribution and a smaller relative RV signal.
Examining the relative PMs from Figure \ref{f:modelcomp} may appear to support this interpretation as the average PMs there are pointing towards the southeast, which would suggest the tidal component is playing a significant role and thus those stars would need to lie in front of the SMC center.
However, whether this is simply an incorrectly modeled, but still smooth, distribution of stars or a more discrete substructure that is sitting in front of the SMC, we are currently unable to constrain.

The second notable area is in the northwest region of the SMC, where there is a clear trend in the RV data of a positive relative RV.
Our model, due to focusing on the central core of the SMC, does not fully cover the sampled area here, but it is clear from the trends in our models that this large, positive RV would not be predicted.
Unlike the southwest region, where it may be a relatively small correction to the distance assumption (as our model does capture some of the expected RV scatter there), the northwest region appears to be a substantial deviation.
Perhaps the most straightforward explanation would be the presence of a stellar substructure that breaks with the apparently smooth RG spatial distributions.
However more work will be required, perhaps with the aid of simulations, in better understanding the type of structure that could lead to this signal in the northwest.

Underscoring all of these comparisons is the clear need for more RV measurements in the SMC.
Already we can see that the RVs can help us in constraining a likely value for $V_0$sin$i$ and in identifying potentially interesting substructure in the SMC.
An expanded catalog will allow us to go further and better map the full kinematic structure of the SMC, including better constraining the LON PA of the rotation plane, which we can see poses a significant challenge, given its complicated interactions with the tidal expansion of the SMC.

Despite these complications, all of our best-fit class of models appear to be converging on the need for both a non-zero tidal expansion component and a non-zero coherent rotation curve in order to explain the observed PMs and RVs, placing this work in contrast to some of the most recent SMC \gaia\ efforts.
Of \cite{deleo20}, \cite{murray19}, and \cite{gaiahelmi18}, only the last study found any need for rotation (see values in Table \ref{tab:modelparam}.
While the preferred inclination angle is similar to our median value, their preferred rotation curve is noticeably slower than ours and their LON PA is nearly opposite to ours.
However, this is not necessarily surprising, given both the differences in the assumed systemic properties (PM, kinematic center, and distance) and that their sample includes both young and old stars, which will have very different kinematic histories.
Thus it becomes difficult to fully assess this as an apples-to-apples comparison.

Similarly, the results of \cite{diteodoro19} and \cite{murray19} are both focused on understanding the kinematics of the gas of the SMC, making a direct comparison challenging.
Given the turbulent interaction history between the SMC and LMC, the dynamics of the SMC gas could have very likely been perturbed beyond a simple rotating disk, and these perturbations would have then transferred to the young stars examined in \cite{murray19}.
That the older, collisionless stars appear to disagree with the inferred kinematics of the gas and the youngest, most recently formed stars in the SMC avoids placing these in conflict as well.

The most direct comparison in the literature to our work is \cite{deleo20} where they also focused on SMC RGs with a combination of RV and PM information.
There they concluded that the magnitude of the tidal disruption occurring prevented any potential rotation from being detected.
Certainly, in looking at the residual PMs on a large scale (like in the top panel of Figure \ref{f:respm_data}), it would not appear that there is any obvious rotation.
However, as we have demonstrated, with a careful removal of viewing perspective effects and simultaneously fitting for both rotation and tidal effects, there is a real rotation signature across the whole of the SMC.
Taken together, these differing conclusions underscore the power of our forward modeling technique and the need for a variety of analytical approaches to fully investigate the complexities of the SMC.

\defcitealias{diteodoro19}{DT19}
\defcitealias{dobbie14a}{D14}
\defcitealias{gaiahelmi18}{GH18}

\begin{table*}
\caption{Rotation Model Parameters.}
    \begin{tabular}{cllrrrrr}
    \tableline
    \tableline
         (1) & Model &  & \citetalias{diteodoro19} & \citetalias{gaiahelmi18} & \citetalias{dobbie14a} & Best-Fit Median & Low Rot. Model \\
         \tableline
         (2) & Distance $D_0$ & kpc & 63 & 62.8 & 60.3 & 60.6 & 60.6 \\
         (3) & Proper motion $\mu_{W}$ & mas yr$^{-1}$ & $-0.77$ & -0.80 & $-0.78$ & $-0.67^{a}$ & $-0.67^{a}$ \\ 
         (4) & Proper motion $\mu_{N}$ & mas yr$^{-1}$ & $-1.12$  & -1.23 & $-1.11$ & $-1.20^{a}$ & $-1.20^{a}$ \\
         (5) & RA $\alpha_0$ (J2000) & deg & 15.24 & 16.26 & 16.25 & 13.04$^{a}$ & 13.04$^{a}$ \\
         (6) & Dec $\delta_0$ (J2000) & deg & $-72.27$ & -72.42 & $-72.42$ & $-73.10^{a}$ & $-73.10^{a}$ \\
         (7) & Systemic Velocity $V_{\mathrm{sys}}$ & km s$^{-1}$  & 148 & 145.6 & 147.8 & 148.0$^{b}$ & 148.0$^{b}$ \\
         (8) & Inclination angle $i$ & deg & 51 & 75.9 & 50 & 80$^{c}$ & 40$^{c}$\\ 
         (9) & Position Angle of LON $\Theta$ & deg & 66 & 5 & 125 & 190$^{c}$ & 170$^{c}$\\ 
         (10) & Rotation Velocity $V_0$ & km s$^{-1}$ & 47 & 3.0$^{d}$ & 30$^{e}$ & 20$^{c}$ & 10$^{c}$\\ 
         (11) & Scale Radius $R_0$ & kpc & 2.8 & $-^{d}$ & $-^{e}$ & 1.0 & 1.0 \\ 
         (12) & Precession/Nutation $di/dt$ & deg Gyr$^{-1}$ & $281$ & $-$ & 140 & $-$ & $-$ \\
         (13) & Tidal Scale $V_{\mathrm{tidal}}$ & km s$^{-1}$ kpc$^{-1}$& $-$ & $-$ & $-$ & 10$^{c}$ & 10$^{c}$\\
         \tableline
    \end{tabular}
    \tablecomments{For \cite{diteodoro19} (DT19 in table), the values listed are sourced from their Table 1. For \cite{dobbie14a} (D14 in table), representative values are sourced from the abstract overview of the examined range of models. For \cite{gaiahelmi18} (GH18 in table), the values listed are sourced from Table B.2 for binned sources inside of 3 degrees. Blank spaces indicate that this value was not explicitly included in the models or in the final reported values. \\
    a. Measured from the \gaia\ RG data. \\
    b. From \cite{deleo20}. \\
    c. Indicates values allowed to vary in model fitting \\
    d. For \citetalias{gaiahelmi18}, the rotation is originally given as a rotational velocity in mas yr$^{-1}$. For a comparison to other models, we have converted this to km s$^{-1}$ kpc$^{-1}$, increasing out to 3 kpc. \\
    e. For \citetalias{dobbie14a}, the rotation velocity noted is in units of km s$^{-1}$ kpc$^{-1}$, increasing out to a few kpc, similar to \citetalias{gaiahelmi18}. }
    \label{tab:modelparam}
\end{table*}

\begin{table*}
\centering
\caption{Best-Fit Median Model Values \& Uncertainties.}
    \begin{tabular}{cllc}
    \tableline
    \tableline
         (1) & Model &  & Median \& MAD\\
         \tableline
         (2) & Distance $D_0$ & kpc  & 60.6$^{a}$ \\
         (3) & Proper motion $\mu_{W}$ & mas yr$^{-1}$  & $-0.67^{b}$ \\ 
         (4) & Proper motion $\mu_{N}$ & mas yr$^{-1}$ & $-1.20^{b}$  \\
         (5) & RA $\alpha_0$ (J2000) & deg & 13.04$^{b}$ \\
         (6) & Dec $\delta_0$ (J2000) & deg & $-73.10^{b}$ \\
         (7) & Systemic Velocity $V_{\mathrm{sys}}$ & km s$^{-1}$ & 148.0$^{c}$ \\
         (8) & Inclination angle $i$ & deg & 80 $\pm\ 9$\\ 
         (9) & Position Angle of LON $\Theta^{d}$ & deg & 190 $\pm\ 42$\\ 
         (10) & Rotation Velocity $V_0$ & km s$^{-1}$ & 20 $\pm\ 7$\\ 
         (11) & Scale Radius $R_0$ & kpc & 1.0 \\ 
         (12) & Precession/Nutation $di/dt$ & deg Gyr$^{-1}$ & 0.0  \\
         (13) & Tidal Scale $V_{\mathrm{tidal}}$ & km s$^{-1}$ kpc$^{-1}$& 10 $\pm\ 1$\\
         \tableline
    \end{tabular}
    \tablecomments{
    a. From \cite{jacyszyn17}. \\
    b. Measured from the \gaia\ RG data. \\
    c. From \cite{deleo20}. \\
    d. Measured east over north.}
    \label{tab:modelerror}
\end{table*}

\section{Conclusions and Discussion}\label{sec:conc}

With the release of \gaia\ DR2, the amount of kinematic information available for the Magellanic system expanded exponentially. 
Beyond PM information, novel spectroscopic surveys of the Magellanic Clouds have added new RV and metallicity information, bringing us closer to a more complete understanding of the Clouds. 
New combinations of these different parameter spaces will offer the opportunity to constrain the Clouds' interactions and underlying physical properties. 
To help set the foundation for these more complex models and analyses of the SMC, we have presented a novel approach to disentangling the PMs of the SMC RG population.

We create a mock SMC which includes a tidal expansion and a rotation component for its stars and use existing frameworks for transforming 3D velocities for resolved stellar systems into observable quantities (\citetalias{vdM02}) to predict the PM distribution of the \gaia\ RGs. 
The result is a model capable of providing a physically intuitive explanation to the otherwise unusual-appearing PM structure of the SMC, one which needs both a coherent rotation and tidal expansion to fully capture the PM behavior. 
Rotation alone is insufficient to explain the observed PM behavior in the SMC, and invoking tidal expansion of the stars, similar to the kinematics observed in simulations of dispersion-supported satellite systems on plunging orbits, offers a possible explanation.
However, to model this effect, we are required to work with the velocities in true 3D space as the 3D structure of the SMC will drastically affect the implied PM distribution due to the tidal expansion and geometric projection effects. Therefore, this forward modeling approach is crucial.

Using the data-derived systemic properties, we find a preference for kinematic models with a moderate rotation curve ($\sim20$ km s$^{-1}$), a LON pointing from the north to the south ($\Theta\sim190$ degrees), a large inclination angle ($\sim80$ degrees), and a tidal scale factor of 10 km s$^{-1}$ kpc$^{-1}$. 
However, there exists a degeneracy between the magnitude of the rotation and the inclination of the rotation plane, with larger inclinations leading to preferring larger rotation velocities.
To break this degeneracy and fully constrain the rotation curve, we will require more RV measurements to provide full 3D motions for a larger portion of our PM catalog.

Already, we have seen the ability of the currently available RVs to place some constraints on the kinematics, with the actual $V_0$sin$i$ likely being smaller than the one in our median model.
However, the slower-rotating, low inclination model tested does struggle to replicate the fast-rising tangential PM component seen in the PMs, suggesting that the reality potentially lies somewhere in the middle of these two models.
Beyond the PM and RV tensions, there appear to be significant deviations in the RV data that will require more sophisticated 3D modeling of the SMC and potentially allowing for spatial substructure to explain some of the more unusual signals, the significance of which will be clarified as the total RVs available continue to increase.
Accordingly, we look towards those future RV surveys and data releases which will allow us to build a more complete picture of the SMC kinematic structure.
In particular, improving RG sampling in the outer regions of the SMC (further than $\sim3$ degrees from the SMC center) will be a crucial constraint on the rotation plane of the SMC, both increasing the number densities in the already observed southwest and northeast outer regions in addition to filling in the sparse southeast and northwest regions.

The combination of the existence of coherent stellar rotation and a clear offset in the systemic RG properties (center, bulk PM) from previously measured studies with a mixed stellar sample suggests the need for a reassessment of how we understand the SMC.
To this, we propose a new interpretation similar to the Bullet Cluster, where in the last interaction with the LMC, the SMC gas was slowed and pulled towards the LMC, while the RG stars passed through collision-free, kinematically separating from the newly forming stars.
Recent work for the LMC has revealed similar population-dependent kinematics, potentially due to interactions with the SMC, providing new constraints on understanding this system \citep{wan20}.
Moving forward, we underscore the necessity of identifying distinct SMC stellar populations and treating their kinematics separately, which may provide a better handle on constraining the LMC-SMC interaction, as opposed to measuring averaged systemic SMC properties over multiple stellar populations.
Future work will be necessary to properly characterize the kinematics of different stellar populations from each other, including modifying existing prescriptions or introducing new ones capable of capturing the behavior younger stars born from turbulent gas.

\acknowledgements
NK is supported by NSF CAREER award 1455260.

This work has made use of data from the European Space Agency (ESA) mission
\gaia\ (\url{https://www.cosmos.esa.int/gaia}), processed by the \gaia\
Data Processing and Analysis Consortium (DPAC,
\url{https://www.cosmos.esa.int/web/gaia/dpac/consortium}). Funding for the DPAC
has been provided by national institutions, in particular the institutions
participating in the \gaia\ Multilateral Agreement.

Funding for the Sloan Digital Sky Survey IV has been provided by the Alfred P. Sloan Foundation, the U.S. Department of Energy Office of Science, and the Participating Institutions. SDSS acknowledges support and resources from the Center for High-Performance Computing at the University of Utah. The SDSS web site is www.sdss.org.

SDSS is managed by the Astrophysical Research Consortium for the Participating Institutions of the SDSS Collaboration including the Brazilian Participation Group, the Carnegie Institution for Science, Carnegie Mellon University, the Chilean Participation Group, the French Participation Group, Harvard-Smithsonian Center for Astrophysics, Instituto de Astrofísica de Canarias, The Johns Hopkins University, Kavli Institute for the Physics and Mathematics of the Universe (IPMU) / University of Tokyo, the Korean Participation Group, Lawrence Berkeley National Laboratory, Leibniz Institut für Astrophysik Potsdam (AIP), Max-Planck-Institut für Astronomie (MPIA Heidelberg), Max-Planck-Institut für Astrophysik (MPA Garching), Max-Planck-Institut für Extraterrestrische Physik (MPE), National Astronomical Observatories of China, New Mexico State University, New York University, University of Notre Dame, Observatório Nacional / MCTI, The Ohio State University, Pennsylvania State University, Shanghai Astronomical Observatory, United Kingdom Participation Group, Universidad Nacional Autónoma de México, University of Arizona, University of Colorado Boulder, University of Oxford, University of Portsmouth, University of Utah, University of Virginia, University of Washington, University of Wisconsin, Vanderbilt University, and Yale University.

\bibliography{Zivick_bib}{}

\begin{thebibliography}{}
\expandafter\ifx\csname natexlab\endcsname\relax\def\natexlab#1{#1}\fi
\providecommand{\url}[1]{\href{#1}{#1}}
\providecommand{\dodoi}[1]{doi:~\href{http://doi.org/#1}{\nolinkurl{#1}}}
\providecommand{\doeprint}[1]{\href{http://ascl.net/#1}{\nolinkurl{http://ascl.net/#1}}}
\providecommand{\doarXiv}[1]{\href{https://arxiv.org/abs/#1}{\nolinkurl{https://arxiv.org/abs/#1}}}

\bibitem[{{Ahumada} {et~al.}(2019){Ahumada}, {Allende Prieto}, {Almeida},
  {Anders}, {Anderson}, {Andrews}, {Anguiano}, {Arcodia}, {Armengaud},
  {Aubert}, {Avila}, {Avila-Reese}, {Badenes}, {Balland}, {Barger},
  {Barrera-Ballesteros}, {Basu}, {Bautista}, {Beaton}, {Beers}, {Benavides},
  {Bender}, {Bernardi}, {Bershady}, {Beutler}, {Moni Bidin}, {Bird}, {Bizyaev},
  {Blanc}, {Blanton}, {Boquien}, {Borissova}, {Bovy}, {Brandt}, {Brinkmann},
  {Brownstein}, {Bundy}, {Bureau}, {Burgasser}, {Burtin}, {Cano-Diaz},
  {Capasso}, {Cappellari}, {Carrera}, {Chabanier}, {Chaplin}, {Chapman},
  {Cherinka}, {Chiappini}, {Choi}, {Chojnowski}, {Chung}, {Clerc}, {Coffey},
  {Comerford}, {Comparat}, {da Costa}, {Cousinou}, {Covey}, {Crane}, {Cunha},
  {da Silva Ilha}, {Dai}, {Damsted}, {Darling}, {Horta Darrington}, {Davidson},
  {Davies}, {Dawson}, {De}, {de la Macorra}, {De Lee}, {Queiroz}, {Deconto
  Machado}, {de la Torre}, {Dell'Agli}, {du Mas des Bourboux},
  {Diamond-Stanic}, {Dillon}, {Donor}, {Drory}, {Duckworth}, {Dwelly},
  {Ebelke}, {Eftekharzadeh}, {Davis Eigenbrot}, {Elsworth}, {Eracleous},
  {Erfanianfar}, {Escoffier}, {Fan}, {Farr}, {Fernandez-Trincado}, {Feuillet},
  {Finoguenov}, {Fofie}, {Fraser-McKelvie}, {Frinchaboy}, {Fromenteau}, {Fu},
  {Galbany}, {Garcia}, {Garcia-Hernandez}, {Garma Oehmichen}, {Ge}, {Geimba
  Maia}, {Geisler}, {Gelfand }, {Goddy}, {Le Goff}, {Gonzalez-Perez},
  {Grabowski}, {Green}, {Grier}, {Guo}, {Guy}, {Harding}, {Hasselquist},
  {Hawken}, {Hayes}, {Hearty}, {Hekker}, {Hogg}, {Holtzman}, {Hou}, {Hsieh},
  {Huber}, {Hunt}, {Ider Chitham}, {Imig}, {Jaber}, {Jimenez Angel}, {Johnson},
  {Jones}, {Jonsson}, {Jullo}, {Kim}, {Kinemuchi}, {Kirkpatrick}, {Kite},
  {Klaene}, {Kneib}, {Kollmeier}, {Kong}, {Kounkel}, {Krishnarao}, {Lacerna},
  {Lan}, {Lane}, {Law}, {Leung}, {Lewis}, {Li}, {Lian}, {Lin}, {Long},
  {Longa-Pena}, {Lundgren}, {Lyke}, {Mackereth}, {MacLeod}, {Majewski},
  {Manchado}, {Maraston}, {Martini}, {Masseron}, {Masters}, {Mathur},
  {McDermid}, {Merloni}, {Merrifield}, {Meszaros}, {Miglio}, {Minniti},
  {Minsley}, {Miyaji}, {Gohar Mohammad}, {Mosser}, {Mueller}, {Muna},
  {Munoz-Gutierrez}, {Myers}, {Nadathur}, {Nair}, {Correa do Nascimento},
  {Nevin}, {Newman}, {Nidever}, {Nitschelm}, {Noterdaeme}, {O'Connell},
  {Olmstead}, {Oravetz}, {Oravetz}, {Osorio}, {Pace}, {Padilla},
  {Palanque-Delabrouille}, {Palicio}, {Pan}, {Pan}, {Parker}, {Paviot},
  {Peirani}, {Pena Ramrez}, {Penny}, {Percival}, {Perez-Fournon},
  {Perez-Rafols}, {Petitjean}, {Pieri}, {Pinsonneault}, {Poovelil}, {Povick},
  {Prakash}, {Price-Whelan}, {Raddick}, {Raichoor}, {Ray}, {Barboza Rembold},
  {Rezaie}, {Riffel}, {Riffel}, {Rix}, {Robin}, {Roman-Lopes}, {Roman-Zuniga},
  {Rose}, {Ross}, {Rossi}, {Rowlands}, {Rubin}, {Salvato}, {Sanchez},
  {Sanchez-Menguiano}, {Sanchez-Gallego}, {Sayres}, {Schaefer}, {Schiavon},
  {Schimoia}, {Schlafly}, {Schlegel}, {Schneider}, {Schultheis}, {Schwope},
  {Seo}, {Serenelli}, {Shafieloo}, {Shamsi}, {Shao}, {Shen}, {Shetrone},
  {Shirley}, {Silva Aguirre}, {Simon}, {Skrutskie}, {Slosar}, {Smethurst},
  {Sobeck}, {Cervantes Sodi}, {Souto}, {Stark}, {Stassun}, {Steinmetz},
  {Stello}, {Stermer}, {Storchi-Bergmann}, {Streblyanska}, {Stringfellow},
  {Stutz}, {Suarez}, {Sun}, {Taghizadeh-Popp}, {Talbot}, {Tayar}, {Thakar},
  {Theriault}, {Thomas}, {Thomas}, {Tinker}, {Tojeiro}, {Hernandez Toledo},
  {Tremonti}, {Troup}, {Tuttle}, {Unda-Sanzana}, {Valentini},
  {Vargas-Gonzalez}, {Vargas-Magana}, {Vazquez-Mata}, {Vivek}, {Wake}, {Wang},
  {Weaver}, {Weijmans}, {Wild}, {Wilson}, {Wilson}, {Wolthuis}, {Wood-Vasey},
  {Yan}, {Yang}, {Yeche}, {Zamora}, {Zarrouk}, {Zasowski}, {Zhang}, {Zhao},
  {Zhao}, {Zheng}, {Zheng}, {Zhu}, \& {Zou}}]{SDSSDR16}
{Ahumada}, R., {Allende Prieto}, C., {Almeida}, A., {et~al.} 2019, arXiv
  e-prints, arXiv:1912.02905.
\newblock \doarXiv{1912.02905}

\bibitem[{{Besla} {et~al.}(2007){Besla}, {Kallivayalil}, {Hernquist},
  {Robertson}, {Cox}, {van der Marel}, \& {Alcock}}]{besla07}
{Besla}, G., {Kallivayalil}, N., {Hernquist}, L., {et~al.} 2007, \apj, 668,
  949, \dodoi{10.1086/521385}

\bibitem[{{Besla} {et~al.}(2012){Besla}, {Kallivayalil}, {Hernquist}, {van der
  Marel}, {Cox}, \& {Kere{\v s}}}]{besla12}
---. 2012, \mnras, 421, 2109, \dodoi{10.1111/j.1365-2966.2012.20466.x}

\bibitem[{{Blanton} {et~al.}(2017){Blanton}, {Bershady}, {Abolfathi},
  {Albareti}, {Allende Prieto}, {Almeida}, {Alonso-Garc{\'\i}a}, {Anders},
  {Anderson}, {Andrews}, {Aquino-Ort{\'\i}z}, {Arag{\'o}n-Salamanca},
  {Argudo-Fern{\'a}ndez}, {Armengaud}, {Aubourg}, {Avila-Reese}, {Badenes},
  {Bailey}, {Barger}, {Barrera-Ballesteros}, {Bartosz}, {Bates}, {Baumgarten},
  {Bautista}, {Beaton}, {Beers}, {Belfiore}, {Bender}, {Berlind}, {Bernardi},
  {Beutler}, {Bird}, {Bizyaev}, {Blanc}, {Blomqvist}, {Bolton}, {Boquien},
  {Borissova}, {van den Bosch}, {Bovy}, {Brandt}, {Brinkmann}, {Brownstein},
  {Bundy}, {Burgasser}, {Burtin}, {Busca}, {Cappellari}, {Delgado Carigi},
  {Carlberg}, {Carnero Rosell}, {Carrera}, {Chanover}, {Cherinka}, {Cheung},
  {G{\'o}mez Maqueo Chew}, {Chiappini}, {Choi}, {Chojnowski}, {Chuang},
  {Chung}, {Cirolini}, {Clerc}, {Cohen}, {Comparat}, {da Costa}, {Cousinou},
  {Covey}, {Crane}, {Croft}, {Cruz-Gonzalez}, {Garrido Cuadra}, {Cunha},
  {Damke}, {Darling}, {Davies}, {Dawson}, {de la Macorra}, {Dell'Agli}, {De
  Lee}, {Delubac}, {Di Mille}, {Diamond-Stanic}, {Cano-D{\'\i}az}, {Donor},
  {Downes}, {Drory}, {du Mas des Bourboux}, {Duckworth}, {Dwelly}, {Dyer},
  {Ebelke}, {Eigenbrot}, {Eisenstein}, {Emsellem}, {Eracleous}, {Escoffier},
  {Evans}, {Fan}, {Fern{\'a}ndez-Alvar}, {Fernandez-Trincado}, {Feuillet},
  {Finoguenov}, {Fleming}, {Font-Ribera}, {Fredrickson}, {Freischlad},
  {Frinchaboy}, {Fuentes}, {Galbany}, {Garcia-Dias},
  {Garc{\'\i}a-Hern{\'a}ndez}, {Gaulme}, {Geisler}, {Gelfand},
  {Gil-Mar{\'\i}n}, {Gillespie}, {Goddard}, {Gonzalez-Perez}, {Grabowski},
  {Green}, {Grier}, {Gunn}, {Guo}, {Guy}, {Hagen}, {Hahn}, {Hall}, {Harding},
  {Hasselquist}, {Hawley}, {Hearty}, {Gonzalez Hern{\'a}ndez}, {Ho}, {Hogg},
  {Holley-Bockelmann}, {Holtzman}, {Holzer}, {Huehnerhoff}, {Hutchinson},
  {Hwang}, {Ibarra-Medel}, {da Silva Ilha}, {Ivans}, {Ivory}, {Jackson},
  {Jensen}, {Johnson}, {Jones}, {J{\"o}nsson}, {Jullo}, {Kamble}, {Kinemuchi},
  {Kirkby}, {Kitaura}, {Klaene}, {Knapp}, {Kneib}, {Kollmeier}, {Lacerna},
  {Lane}, {Lang}, {Law}, {Lazarz}, {Lee}, {Le Goff}, {Liang}, {Li}, {Li},
  {Lian}, {Lima}, {Lin}, {Lin}, {Bertran de Lis}, {Liu}, {de Icaza Lizaola},
  {Long}, {Lucatello}, {Lundgren}, {MacDonald}, {Deconto Machado}, {MacLeod},
  {Mahadevan}, {Geimba Maia}, {Maiolino}, {Majewski}, {Malanushenko},
  {Malanushenko}, {Manchado}, {Mao}, {Maraston}, {Marques-Chaves}, {Masseron},
  {Masters}, {McBride}, {McDermid}, {McGrath}, {McGreer}, {Medina Pe{\~n}a},
  {Melendez}, {Merloni}, {Merrifield}, {Meszaros}, {Meza}, {Minchev},
  {Minniti}, {Miyaji}, {More}, {Mulchaey}, {M{\"u}ller-S{\'a}nchez}, {Muna},
  {Munoz}, {Myers}, {Nair}, {Nandra}, {Correa do Nascimento}, {Negrete},
  {Ness}, {Newman}, {Nichol}, {Nidever}, {Nitschelm}, {Ntelis}, {O'Connell},
  {Oelkers}, {Oravetz}, {Oravetz}, {Pace}, {Padilla}, {Palanque-Delabrouille},
  {Alonso Palicio}, {Pan}, {Parejko}, {Parikh}, {P{\^a}ris}, {Park}, {Patten},
  {Peirani}, {Pellejero-Ibanez}, {Penny}, {Percival}, {Perez-Fournon},
  {Petitjean}, {Pieri}, {Pinsonneault}, {Pisani}, {Poleski}, {Prada},
  {Prakash}, {Queiroz}, {Raddick}, {Raichoor}, {Barboza Rembold}, {Richstein},
  {Riffel}, {Riffel}, {Rix}, {Robin}, {Rockosi}, {Rodr{\'\i}guez-Torres},
  {Roman-Lopes}, {Rom{\'a}n-Z{\'u}{\~n}iga}, {Rosado}, {Ross}, {Rossi}, {Ruan},
  {Ruggeri}, {Rykoff}, {Salazar-Albornoz}, {Salvato}, {S{\'a}nchez}, {Aguado},
  {S{\'a}nchez-Gallego}, {Santana}, {Santiago}, {Sayres}, {Schiavon}, {da Silva
  Schimoia}, {Schlafly}, {Schlegel}, {Schneider}, {Schultheis}, {Schuster},
  {Schwope}, {Seo}, {Shao}, {Shen}, {Shetrone}, {Shull}, {Simon}, {Skinner},
  {Skrutskie}, {Slosar}, {Smith}, {Sobeck}, {Sobreira}, {Somers}, {Souto},
  {Stark}, {Stassun}, {Stauffer}, {Steinmetz}, {Storchi-Bergmann},
  {Streblyanska}, {Stringfellow}, {Su{\'a}rez}, {Sun}, {Suzuki}, {Szigeti},
  {Taghizadeh-Popp}, {Tang}, {Tao}, {Tayar}, {Tembe}, {Teske}, {Thakar},
  {Thomas}, {Thompson}, {Tinker}, {Tissera}, {Tojeiro}, {Hernandez Toledo}, {de
  la Torre}, {Tremonti}, {Troup}, {Valenzuela}, {Martinez Valpuesta},
  {Vargas-Gonz{\'a}lez}, {Vargas-Maga{\~n}a}, {Vazquez}, {Villanova}, {Vivek},
  {Vogt}, {Wake}, {Walterbos}, {Wang}, {Weaver}, {Weijmans}, {Weinberg},
  {Westfall}, {Whelan}, {Wild}, {Wilson}, {Wood-Vasey}, {Wylezalek}, {Xiao},
  {Yan}, {Yang}, {Ybarra}, {Y{\`e}che}, {Zakamska}, {Zamora}, {Zarrouk},
  {Zasowski}, {Zhang}, {Zhao}, {Zheng}, {Zheng}, {Zhou}, {Zhou}, {Zhu},
  {Zoccali}, \& {Zou}}]{Blanton17}
{Blanton}, M.~R., {Bershady}, M.~A., {Abolfathi}, B., {et~al.} 2017, \aj, 154,
  28, \dodoi{10.3847/1538-3881/aa7567}

\bibitem[{{De Leo} {et~al.}(2020){De Leo}, {Carrera}, {Noel}, {Read}, {Erkal},
  \& {Gallart}}]{deleo20}
{De Leo}, M., {Carrera}, R., {Noel}, N. E.~D., {et~al.} 2020, arXiv e-prints,
  arXiv:2002.11138.
\newblock \doarXiv{2002.11138}

\bibitem[{{Di Teodoro} {et~al.}(2019){Di Teodoro}, {McClure-Griffiths},
  {Jameson}, {D{\'e}nes}, {Dickey}, {Stanimirovi{\'c}}, {Staveley-Smith},
  {Anderson}, {Bunton}, {Chippendale}, {Lee-Waddell}, {MacLeod}, \&
  {Voronkov}}]{diteodoro19}
{Di Teodoro}, E.~M., {McClure-Griffiths}, N.~M., {Jameson}, K.~E., {et~al.}
  2019, \mnras, 483, 392, \dodoi{10.1093/mnras/sty3095}

\bibitem[{{Dobbie} {et~al.}(2014){Dobbie}, {Cole}, {Subramaniam}, \&
  {Keller}}]{dobbie14a}
{Dobbie}, P.~D., {Cole}, A.~A., {Subramaniam}, A., \& {Keller}, S. 2014,
  \mnras, 442, 1663, \dodoi{10.1093/mnras/stu910}

\bibitem[{{Evans} \& {Howarth}(2008)}]{evans08}
{Evans}, C.~J., \& {Howarth}, I.~D. 2008, \mnras, 386, 826,
  \dodoi{10.1111/j.1365-2966.2008.13012.x}

\bibitem[{{Freedman} \& {Diaconis}(1981)}]{FD81}
{Freedman}, D., \& {Diaconis}, P.~Z. 1981, Wahrscheinlichkeitstheorie verw
  Gebiete, 57, \dodoi{10.1007/BF01025868}

\bibitem[{{Gaia Collaboration} {et~al.}(2016{\natexlab{a}}){Gaia
  Collaboration}, {Brown}, {Vallenari}, {Prusti}, {de Bruijne}, {Mignard},
  {Drimmel}, {Babusiaux}, {Bailer-Jones}, {Bastian}, \& et~al.}]{gaiadr1a}
{Gaia Collaboration}, {Brown}, A.~G.~A., {Vallenari}, A., {et~al.}
  2016{\natexlab{a}}, \aap, 595, A2, \dodoi{10.1051/0004-6361/201629512}

\bibitem[{{Gaia Collaboration} {et~al.}(2016{\natexlab{b}}){Gaia
  Collaboration}, {Prusti}, {de Bruijne}, {Brown}, {Vallenari}, {Babusiaux},
  {Bailer-Jones}, {Bastian}, {Biermann}, {Evans}, \& et~al.}]{gaiadr1b}
{Gaia Collaboration}, {Prusti}, T., {de Bruijne}, J.~H.~J., {et~al.}
  2016{\natexlab{b}}, \aap, 595, A1, \dodoi{10.1051/0004-6361/201629272}

\bibitem[{{Gaia Collaboration} {et~al.}(2018{\natexlab{a}}){Gaia
  Collaboration}, {Brown}, {Vallenari}, {Prusti}, {de Bruijne}, {Babusiaux},
  {Bailer-Jones}, {Biermann}, {Evans}, {Eyer}, \& et~al.}]{gaiadr218b}
{Gaia Collaboration}, {Brown}, A.~G.~A., {Vallenari}, A., {et~al.}
  2018{\natexlab{a}}, \aap, 616, A1, \dodoi{10.1051/0004-6361/201833051}

\bibitem[{{Gaia Collaboration} {et~al.}(2018{\natexlab{b}}){Gaia
  Collaboration}, {Helmi}, {van Leeuwen}, {McMillan}, {Massari}, {Antoja},
  {Robin}, {Lindegren}, {Bastian}, {Arenou}, \& et~al.}]{gaiahelmi18}
{Gaia Collaboration}, {Helmi}, A., {van Leeuwen}, F., {et~al.}
  2018{\natexlab{b}}, \aap, 616, A12, \dodoi{10.1051/0004-6361/201832698}

\bibitem[{{Gaia Collaboration} {et~al.}(2018{\natexlab{c}}){Gaia
  Collaboration}, {Babusiaux}, {van Leeuwen}, {Barstow}, {Jordi}, {Vallenari},
  {Bossini}, {Bressan}, {Cantat-Gaudin}, {van Leeuwen}, \& et~al.}]{gaiadr2hrd}
{Gaia Collaboration}, {Babusiaux}, C., {van Leeuwen}, F., {et~al.}
  2018{\natexlab{c}}, \aap, 616, A10, \dodoi{10.1051/0004-6361/201832843}

\bibitem[{{Grady} {et~al.}(2020){Grady}, {Belokurov}, \& {Evans}}]{grady20}
{Grady}, J., {Belokurov}, V., \& {Evans}, N.~W. 2020, arXiv e-prints,
  arXiv:2010.05956.
\newblock \doarXiv{2010.05956}

\bibitem[{{Harris} \& {Zaritsky}(2006)}]{harris06}
{Harris}, J., \& {Zaritsky}, D. 2006, \aj, 131, 2514, \dodoi{10.1086/500974}

\bibitem[{{H{\o}g} {et~al.}(2000){H{\o}g}, {Fabricius}, {Makarov}, {Urban},
  {Corbin}, {Wycoff}, {Bastian}, {Schwekendiek}, \& {Wicenec}}]{hoeg00}
{H{\o}g}, E., {Fabricius}, C., {Makarov}, V.~V., {et~al.} 2000, \aap, 355, L27

\bibitem[{{Jacyszyn-Dobrzeniecka} {et~al.}(2017){Jacyszyn-Dobrzeniecka},
  {Skowron}, {Mr{\'o}z}, {Soszy{\'n}ski}, {Udalski}, {Pietrukowicz}, {Skowron},
  {Poleski}, {Koz{\l}owski}, {Wyrzykowski}, {Pawlak}, {Szyma{\'n}ski}, \&
  {Ulaczyk}}]{jacyszyn17}
{Jacyszyn-Dobrzeniecka}, A.~M., {Skowron}, D.~M., {Mr{\'o}z}, P., {et~al.}
  2017, \actaa, 67, 1, \dodoi{10.32023/0001-5237/67.1.1}

\bibitem[{{Joshi} \& {Panchal}(2019)}]{joshi19}
{Joshi}, Y.~C., \& {Panchal}, A. 2019, \aap, 628, A51,
  \dodoi{10.1051/0004-6361/201834574}

\bibitem[{{Kallivayalil} {et~al.}(2006{\natexlab{a}}){Kallivayalil}, {van der
  Marel}, \& {Alcock}}]{NK06b}
{Kallivayalil}, N., {van der Marel}, R.~P., \& {Alcock}, C. 2006{\natexlab{a}},
  \apj, 652, 1213, \dodoi{10.1086/508014}

\bibitem[{{Kallivayalil} {et~al.}(2006{\natexlab{b}}){Kallivayalil}, {van der
  Marel}, {Alcock}, {Axelrod}, {Cook}, {Drake}, \& {Geha}}]{NK06a}
{Kallivayalil}, N., {van der Marel}, R.~P., {Alcock}, C., {et~al.}
  2006{\natexlab{b}}, \apj, 638, 772, \dodoi{10.1086/498972}

\bibitem[{{Kallivayalil} {et~al.}(2013){Kallivayalil}, {van der Marel},
  {Besla}, {Anderson}, \& {Alcock}}]{NK13}
{Kallivayalil}, N., {van der Marel}, R.~P., {Besla}, G., {Anderson}, J., \&
  {Alcock}, C. 2013, \apj, 764, 161, \dodoi{10.1088/0004-637X/764/2/161}

\bibitem[{{Lamb} {et~al.}(2016){Lamb}, {Oey}, {Segura-Cox}, {Graus}, {Kiminki},
  {Golden-Marx}, \& {Parker}}]{lamb16}
{Lamb}, J.~B., {Oey}, M.~S., {Segura-Cox}, D.~M., {et~al.} 2016, \apj, 817,
  113, \dodoi{10.3847/0004-637X/817/2/113}

\bibitem[{{Lindegren} {et~al.}(2016){Lindegren}, {Lammers}, {Bastian},
  {Hern{\'a}ndez}, {Klioner}, {Hobbs}, {Bombrun}, {Michalik}, {Ramos-Lerate},
  {Butkevich}, {Comoretto}, {Joliet}, {Holl}, {Hutton}, {Parsons},
  {Steidelm{\"u}ller}, {Abbas}, {Altmann}, {Andrei}, {Anton}, {Bach},
  {Barache}, {Becciani}, {Berthier}, {Bianchi}, {Biermann}, {Bouquillon},
  {Bourda}, {Br{\"u}semeister}, {Bucciarelli}, {Busonero}, {Carlucci},
  {Casta{\~n}eda}, {Charlot}, {Clotet}, {Crosta}, {Davidson}, {de Felice},
  {Drimmel}, {Fabricius}, {Fienga}, {Figueras}, {Fraile}, {Gai}, {Garralda},
  {Geyer}, {Gonz{\'a}lez-Vidal}, {Guerra}, {Hambly}, {Hauser}, {Jordan},
  {Lattanzi}, {Lenhardt}, {Liao}, {L{\"o}ffler}, {McMillan}, {Mignard}, {Mora},
  {Morbidelli}, {Portell}, {Riva}, {Sarasso}, {Serraller}, {Siddiqui}, {Smart},
  {Spagna}, {Stampa}, {Steele}, {Taris}, {Torra}, {van Reeven}, {Vecchiato},
  {Zschocke}, {de Bruijne}, {Gracia}, {Raison}, {Lister}, {Marchant},
  {Messineo}, {Soffel}, {Osorio}, {de Torres}, \& {O'Mullane}}]{lindegren16}
{Lindegren}, L., {Lammers}, U., {Bastian}, U., {et~al.} 2016, \aap, 595, A4,
  \dodoi{10.1051/0004-6361/201628714}

\bibitem[{{Majewski} {et~al.}(2017){Majewski}, {Schiavon}, {Frinchaboy},
  {Allende Prieto}, {Barkhouser}, {Bizyaev}, {Blank}, {Brunner}, {Burton},
  {Carrera}, {Chojnowski}, {Cunha}, {Epstein}, {Fitzgerald}, {Garc{\'\i}a
  P{\'e}rez}, {Hearty}, {Henderson}, {Holtzman}, {Johnson}, {Lam}, {Lawler},
  {Maseman}, {M{\'e}sz{\'a}ros}, {Nelson}, {Nguyen}, {Nidever}, {Pinsonneault},
  {Shetrone}, {Smee}, {Smith}, {Stolberg}, {Skrutskie}, {Walker}, {Wilson},
  {Zasowski}, {Anders}, {Basu}, {Beland}, {Blanton}, {Bovy}, {Brownstein},
  {Carlberg}, {Chaplin}, {Chiappini}, {Eisenstein}, {Elsworth}, {Feuillet},
  {Fleming}, {Galbraith-Frew}, {Garc{\'\i}a}, {Garc{\'\i}a-Hern{\'a}ndez},
  {Gillespie}, {Girardi}, {Gunn}, {Hasselquist}, {Hayden}, {Hekker}, {Ivans},
  {Kinemuchi}, {Klaene}, {Mahadevan}, {Mathur}, {Mosser}, {Muna}, {Munn},
  {Nichol}, {O'Connell}, {Parejko}, {Robin}, {Rocha-Pinto}, {Schultheis},
  {Serenelli}, {Shane}, {Silva Aguirre}, {Sobeck}, {Thompson}, {Troup},
  {Weinberg}, \& {Zamora}}]{Majewski17}
{Majewski}, S.~R., {Schiavon}, R.~P., {Frinchaboy}, P.~M., {et~al.} 2017, \aj,
  154, 94, \dodoi{10.3847/1538-3881/aa784d}

\bibitem[{{Martinez-Delgado} {et~al.}(2019){Martinez-Delgado}, {Vivas},
  {Grebel}, {Gallart}, {Pieres}, {Bell}, {Zivick}, {Lemasle}, {Johnson},
  {Carballo-Bello}, {Noel}, {Cioni}, {Choi}, {Besla}, {Schmidt}, {Zaritsky},
  {Gruendl}, {Seibert}, {Nidever}, {Monteagudo}, {Monelli}, {Hubl}, {van der
  Marel}, {Ballesteros}, {Stringfellow}, {Walker}, {Blum}, {Bell}, {Conn},
  {Olsen}, {Martin}, {Chu}, {Inno}, {Boer}, {Kallivayalil}, {De Leo},
  {Beletsky}, \& {Munoz}}]{martinez-delgado19}
{Martinez-Delgado}, D., {Vivas}, A.~K., {Grebel}, E.~K., {et~al.} 2019, arXiv
  e-prints, arXiv:1907.02264.
\newblock \doarXiv{1907.02264}

\bibitem[{{Murray} {et~al.}(2019){Murray}, {Peek}, {Di Teodoro},
  {McClure-Griffiths}, {Dickey}, \& {D{\'e}nes}}]{murray19}
{Murray}, C.~E., {Peek}, J.~E.~G., {Di Teodoro}, E.~M., {et~al.} 2019, \apj,
  887, 267, \dodoi{10.3847/1538-4357/ab510f}

\bibitem[{{Olsen} {et~al.}(2011){Olsen}, {Zaritsky}, {Blum}, {Boyer}, \&
  {Gordon}}]{olsen11}
{Olsen}, K.~A.~G., {Zaritsky}, D., {Blum}, R.~D., {Boyer}, M.~L., \& {Gordon},
  K.~D. 2011, \apj, 737, 29, \dodoi{10.1088/0004-637X/737/1/29}

\bibitem[{{Pe{\~n}arrubia} {et~al.}(2009){Pe{\~n}arrubia}, {Navarro},
  {McConnachie}, \& {Martin}}]{penarrubia09}
{Pe{\~n}arrubia}, J., {Navarro}, J.~F., {McConnachie}, A.~W., \& {Martin},
  N.~F. 2009, \apj, 698, 222, \dodoi{10.1088/0004-637X/698/1/222}

\bibitem[{{Piatek} {et~al.}(2008){Piatek}, {Pryor}, \& {Olszewski}}]{piatek08}
{Piatek}, S., {Pryor}, C., \& {Olszewski}, E.~W. 2008, \aj, 135, 1024,
  \dodoi{10.1088/0004-6256/135/3/1024}

\bibitem[{{Platais} {et~al.}(2015){Platais}, {van der Marel}, {Lennon},
  {Anderson}, {Bellini}, {Sabbi}, {Sana}, \& {Bedin}}]{platais15}
{Platais}, I., {van der Marel}, R.~P., {Lennon}, D.~J., {et~al.} 2015, \aj,
  150, 89, \dodoi{10.1088/0004-6256/150/3/89}

\bibitem[{{Stanimirovi{\'c}} {et~al.}(2004){Stanimirovi{\'c}},
  {Staveley-Smith}, \& {Jones}}]{stanimirovic}
{Stanimirovi{\'c}}, S., {Staveley-Smith}, L., \& {Jones}, P.~A. 2004, \apj,
  604, 176, \dodoi{10.1086/381869}

\bibitem[{{Subramanian} \& {Subramaniam}(2012)}]{ss_sa12}
{Subramanian}, S., \& {Subramaniam}, A. 2012, \apj, 744, 128,
  \dodoi{10.1088/0004-637X/744/2/128}

\bibitem[{{van der Marel} {et~al.}(2002){van der Marel}, {Alves}, {Hardy}, \&
  {Suntzeff}}]{vdM02}
{van der Marel}, R.~P., {Alves}, D.~R., {Hardy}, E., \& {Suntzeff}, N.~B. 2002,
  \aj, 124, 2639, \dodoi{10.1086/343775}

\bibitem[{{van der Marel} \& {Kallivayalil}(2014)}]{vdM14}
{van der Marel}, R.~P., \& {Kallivayalil}, N. 2014, \apj, 781, 121,
  \dodoi{10.1088/0004-637X/781/2/121}

\bibitem[{{van der Marel} \& {Sahlmann}(2016)}]{vdM16}
{van der Marel}, R.~P., \& {Sahlmann}, J. 2016, \apjl, 832, L23,
  \dodoi{10.3847/2041-8205/832/2/L23}

\bibitem[{{Wan} {et~al.}(2020){Wan}, {Guglielmo}, {Lewis}, {Mackey}, \&
  {Ibata}}]{wan20}
{Wan}, Z., {Guglielmo}, M., {Lewis}, G.~F., {Mackey}, D., \& {Ibata}, R.~A.
  2020, \mnras, 492, 782, \dodoi{10.1093/mnras/stz3493}

\bibitem[{{Wilson} {et~al.}(2019){Wilson}, {Hearty}, {Skrutskie}, {Majewski},
  {Holtzman}, {Eisenstein}, {Gunn}, {Blank}, {Henderson}, {Smee}, {Nelson},
  {Nidever}, {Arns}, {Barkhouser}, {Barr}, {Beland}, {Bershady}, {Blanton},
  {Brunner}, {Burton}, {Carey}, {Carr}, {Colque}, {Crane}, {Damke}, {Davidson},
  {Dean}, {Di Mille}, {Don}, {Ebelke}, {Evans}, {Fitzgerald}, {Gillespie},
  {Hall}, {Harding}, {Harding}, {Hammond}, {Hancock}, {Harrison}, {Hope},
  {Horne}, {Karakla}, {Lam}, {Leger}, {MacDonald}, {Maseman}, {Matsunari},
  {Melton}, {Mitcheltree}, {O'Brien}, {O'Connell}, {Patten}, {Richardson},
  {Rieke}, {Rieke}, {Roman-Lopes}, {Schiavon}, {Sobeck}, {Stolberg}, {Stoll},
  {Tembe}, {Trujillo}, {Uomoto}, {Vernieri}, {Walker}, {Weinberg}, {Young},
  {Anthony-Brumfield}, {Bizyaev}, {Breslauer}, {De Lee}, {Downey}, {Halverson},
  {Huehnerhoff}, {Klaene}, {Leon}, {Long}, {Mahadevan}, {Malanushenko},
  {Nguyen}, {Owen}, {S{\'a}nchez-Gallego}, {Sayres}, {Shane}, {Shectman},
  {Shetrone}, {Skinner}, {Stauffer}, \& {Zhao}}]{Wilson19}
{Wilson}, J.~C., {Hearty}, F.~R., {Skrutskie}, M.~F., {et~al.} 2019, \pasp,
  131, 055001, \dodoi{10.1088/1538-3873/ab0075}

\bibitem[{{Zivick} {et~al.}(2018){Zivick}, {Kallivayalil}, {van der Marel},
  {Besla}, {Linden}, {Koz{\l}owski}, {Fritz}, {Kochanek}, {Anderson}, {Sohn},
  {Geha}, \& {Alcock}}]{zivick18}
{Zivick}, P., {Kallivayalil}, N., {van der Marel}, R.~P., {et~al.} 2018, \apj,
  864, 55, \dodoi{10.3847/1538-4357/aad4b0}

\bibitem[{{Zivick} {et~al.}(2019){Zivick}, {Kallivayalil}, {Besla}, {Sohn},
  {van der Marel}, {del Pino}, {Linden}, {Fritz}, \& {Anderson}}]{zivick19}
{Zivick}, P., {Kallivayalil}, N., {Besla}, G., {et~al.} 2019, \apj, 874, 78,
  \dodoi{10.3847/1538-4357/ab0554}

\end{thebibliography}

\end{document}